# Key Technologies for Networked Virtual Environments

**Juan González (juagons4@epsg.upv.es; ORCID: 0000-0002-1005-6038)**

**Fernando Boronat (fboronat@dcom.upv.es; ORCID: 0000-0001-5525-3441)**

**Almanzor Sapena (alsapie@mat.upv.es; ORCID: 0000-0001-8473-6063)**

**Javier Pastor (fjpastor@dib.upv.es; ORCID: 0000-0001-6693-8489)**

* All authors are with the Immersive Interactive Media (IIM) R&D Group, at Universitat Politècnica de València (UPV), Campus de Gandia, España.

Abstract— Thanks to the improvements experienced in technology in the last few years, most especially in virtual reality systems, the number and potential of networked virtual environments or NVEs and their users are increasing. NVEs aim to give distributed users a feeling of immersion in a virtual world and the possibility of interacting with other users or with virtual objects inside it, like when they interact in the real world. Being able to provide that feeling and natural interactions when the users are geographically separated is one of the goals of these systems. Nevertheless, this goal is especially sensitive to different issues, such as different connections with heterogeneous throughput or different network latencies, which can lead to consistency and synchronization problems and, thus, to a worsening of the users' quality of experience or QoE. With the purpose of solving these issues, researchers have proposed and evaluated numerous technical solutions, in fields like network architectures, data distribution and filtering, resource balancing, computing models, predictive modeling and synchronization in NVEs. This paper gathers and classifies them, summarizing their advantages and disadvantages, using a new way of classification. With the current increase of the number of NVEs and the multiple solutions proposed so far, this work aims to become a useful tool and a starting point not only for future researchers in this field but also for those who are new in NVEs development, in which guaranteeing a good users' QoE is essential.

Index Terms—Computing Models, Data Distribution, Data Filtering, Networked Virtual Environment, Predictive Modeling, Resource Balancing, Synchronization.

# 1. Introduction

A Networked Virtual Environment (NVE) is a distributed computer-based system simulating a virtual environment (VE), which supports multiple interconnected remote users who can interact in real time within it in both ways: with the virtual objects in it (e.g., grabbing them) and with the other users in the NVE [1] (e.g. speaking or chatting with them). This type of systems serves as a medium to abstract information and communication for the users. So, with these environments, users can perceive an enriched and immersive experience, while interacting with people even located in different geographic places. Therefore, NVE designers and developers must make them interactive, offering a proper quality of experience (QoE) to the users, based on the NVE desired purpose or application.

The distinct kinds of environments and contents make NVEs suitable for multiple applications. While the main application of NVEs has been in entertainment (e.g., videogames), NVEs have shown a clear potential for other areas as well, such as remote virtual meetings, collaboration, or teaching/learning. There are five main scenarios of NVEs depending on their purpose [2]–[9]: Multiplayer online games (MOGs), educational or training NVEs, NVEs for collaborative work, commercially oriented NVEs, and NVEs for social interaction.

MOGs consist of computer-based games where several remote users play, at the same time, in the same virtual environment. These games can have a real-time play or a turn-based one, which might influence the networking requirements [10]. Some examples are MiMaze [11], City of Heroes (CoH) [12], Kingspray [13], Rokkatan [14], and World of Warcraft (WoW) [15].

Some NVEs can be also used for teaching/learning in different fields, such as in military [16] or medical [17] applications. They can take the trainees away from hazards of real-world training, help with motor learning or enable e-learning. Some examples are Future Visual [18], FishWorld [19], CoMove [20], Virtway [21], and Acadicus [22].

NVEs for collaborative work are virtual environments where remote users communicate, interact, and work together to accomplish a mission or a task [23]. NPSNET-V [24], VSculpt [25], RING [26], MASSIVE [27], DIVE [28], ShareX3D [29], CAVRNUS [30], Spatial [31], Co-Surgeon [32], The Wild [33], and COLLAVIZ [34] are some examples.

Commercially oriented NVEs are used by companies for marketing goals, like researching, designing, testing, or advertising products [35]. Some solutions are found in Spinview [36], Trezi [37], Vizible [38], and Theia Interactive [39].

NVEs for social interaction offer a new way of communication and social interaction for relatives, friends, and strangers to connect from any place around the world and engage in social activities. Second Life [40], IMVU [41],

Diamond Park [42], Virtual Real Meeting [43], Mozilla Hubs [44], CAVRN [45], Decentraland [46], VRChat [47], and Bigscreen [48] are some few examples.

To interact in an immersive way within the NVEs, users usually employ an avatar[1], which is a virtual representation of themselves inside the VE, controlled by one or more input devices (e.g., a mouse, keyboard, joystick, gamepad or haptic) connected to their computers [1], [49]. Furthermore, by giving feedback (e.g., through sensorial effects, such as sounds, vibration, pressure, wind, or aromas) to the events within the virtual world, the users' feeling of immersion is highly improved. Moreover, for visualizing the 3D environments, traditional screens, and Cave Assisted Virtual Environments (CAVEs) [50] can be employed, but nowadays head-mounted displays (HMDs) are gaining momentum, as they are becoming more affordable and are the ones which really provide users with a complete feeling of immersion. Figure 1 shows an overview of the involved structure of a simple NVE with two remote users (A and B) controlling their avatars that can interact with the objects in the VE and between them.

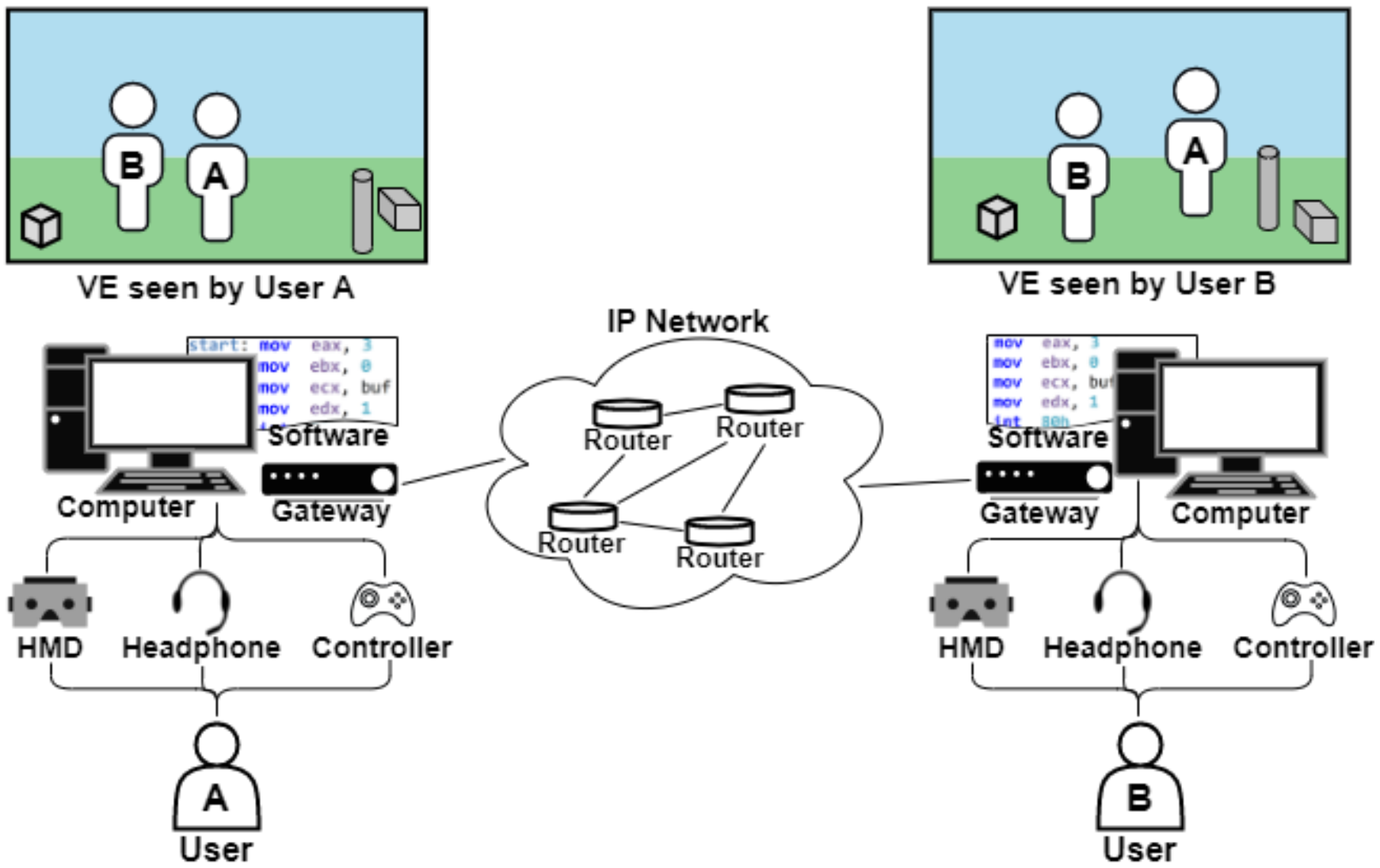

Figure 1. Overview of an NVE with two users (A and B) interacting with objects in the VE and between them.

Before exploring NVEs in deep, it is important to know the base elements that define a VE and how they work. Then the paper will focus on Networked VEs.

In general, VEs comprise both hardware (HW) and software (SW) parts. The former part usually includes computers (with processing units, storage mediums…), displays (normal displays or embedded in HMD), and input/output devices; while the latter mainly consists of digital databases (DBs), media data, operating systems (OS), file systems, and application programs.

On the one hand, some elements in the HW part (e.g., computers) allow to run the SW of the VE and to interact with other HW elements (e.g., input/output devices). The required elements depend on the purpose and complexity of each VE. For instance, a simple 2D computer-based game would just need a computer, a screen, and a keyboard, while an immersive one (e.g., a VR game) may additionally involve a head-mounted display (HMD) and a joystick controller or a haptic device. The input peripherals (e.g., mouse, keyboard or joystick) and output peripherals (e.g., screens, speakers or sensorial effects generators) allow the users to interact with the VE and to receive feedback from it, respectively.

On the other hand, the SW part contributes to simulate the VE. It, provides the structure and logic for running the VE, taking advantage of the HW part for generating and managing it. So, HW and SW parts work together to create a VE and enable users to interact with it.

As HW and SW technology improve, the power, availability, and performance of computers and other devices also increase. These computing advancements lead to VEs with higher quality and realism than long before. Moreover, advanced VEs include 3D-rendered virtual worlds and may require the use of VR equipment, like HMDs and haptic devices, which bring telepresence, or virtual presence at a new level [51]. Consequently, the most advanced VEs need high-performance HW and SW and their development bears greater challenges.

When considering the contents that a VE emulates, there exist four classes of virtual elements: zones, entities, states, and behaviors. *Zones*, also called regions or areas, are virtual geographic areas that divide the virtual environment.

[1] Currently, there is a research effort about using Holograms (e.g., in H2020 EU project VR Together, https://vrtogether.eu)) and it is expected that very soon the avatars will be replaced by holograms in some types of NVEs, e.g., in virtual meetings.

*Entities* are the virtual objects, situated inside the VE. For example, two different cars in a driving simulation are two different entities. Avatars are also a specific kind of entity, representing a user. Users can interact with all the entities in the VE by controlling their avatars. The *States* of an NVE are determined by the values of variables/parameters (or combinations of them) that define the features of the entities or the environment in a specific instant of time. Examples of states would be the position, speed, color, and size of the entities in the VE. *Behaviors* are the actions, events, or conducts that can take place in the NVE (i.e., included in the programmed logic of the NVE). Behaviors allow the NVE logic to take course. For example, a virtual car will suddenly slow down when hitting the brakes.

Regarding the user distribution and communication issues, VEs can allow either one user or multiple ones to interact within it, being single-user or multi-user VEs, respectively. In the latter case, the users can either be in the same place (e.g., PC games allowing two local players, using control devices connected to the same computer) or in separate locations (e.g., networked online games). This latter case of VEs is referred to as *Networked Virtual Environment* (NVE), which allows multiple users simultaneously sharing the VE, even from geographically distant locations, through communication networks. Suitable communication (wired or wireless) networks are needed to enable multiple users to seamlessly interact within the same VE. However, the use of communication networks adds another layer of complexity when designing NVEs, as explained later.

As NVEs combine HW and SW parts for diverse networking tasks, it is of importance to emphasize how those elements intervene.

SW elements (in computers and other HW devices) process the interchanged messages through the communications network. Databases manage the data and interchanged messages (e.g., storing, changing, or removing them), allowing their storage and access by multiple distributed users over the network. Other SW elements, in turn, execute the algorithms that process the data, manage the interactions (e.g., a user pressing a button), and provide feedback (e.g., transmitting a message through the network). The network connections allow the distributed computers or HW devices to interchange messages through wired or wireless links, enabling interaction between separated users.

Considering the relevance of elements of an NVE in tasks associated with networking aspects, the following ones can be emphasized: Nodes (Servers, Clients or Peers), and NVE update messages interchanged between them [5]. *Nodes* are computers or other HW devices connected to a network, enabling communication, or using it. A node could be an individual HW device (e.g., a router) or a group of HW devices (e.g., a datacenter acting as a server). *Servers* are the nodes that distribute resources and information to the clients of an NVE. *Clients* are the nodes that allow users to take part in NVEs (e.g., a computer, game console or a smartphone). Servers connect the users' clients so they can mutually interact. *Peers* are nodes that work as both servers and clients. Peers are used in (serverless) decentralized networks, when there is no central authority or servers, so the server role lies on the clients (peers). Update messages (also referred to as updates) contain information about events happening in the NVE, such as changes of elements of the VE (e.g., change of position of an entity or avatar) due to users' interactions.

Figure 2 shows an example of the architecture of a client/server-based NVE including some of the elements mentioned above with their relationships. Notice that it is only an example and some of the elements could not be present in an NVE, such as NVEs with clients without local DBs, which share the Server DB, or NVEs without server nodes (only peers).

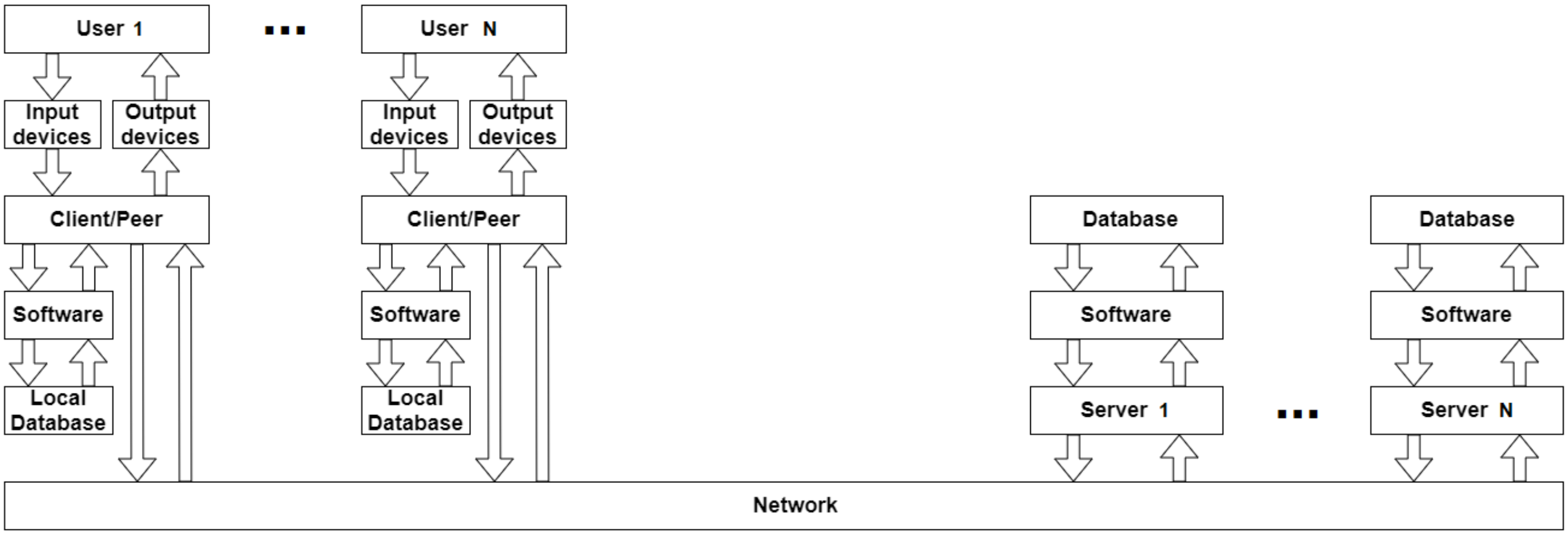

Figure 2. Example of NVE client/server-based architecture.

It is important to note that, as NVEs are complex, developers must delve into the different fields they involve (e.g., networking, computing, database management or 3D modeling). This often requires a diverse team (e.g., network engineers, programmers or database administrators), and multiple tools (e.g., computer graphics software or database management systems). Hence, the design, development, and implementation of NVEs ends up being a laborious task. Additionally, the more fields NVEs encompass, the more opportunities for problems and errors to arise. Two examples are the network-related problems (like network delay) due to the usage of networks, and data inconsistency due to the concurrent access to databases.

In this paper, an NVE is considered as a combination of several components and subcomponents: Network Architecture, Information Management (including Data Filtering and Distribution), Resource Balancing, Time Management (including Predictive Modeling and Synchronization), and Computing Models. The NVE Network Architecture specifies how the nodes or physical devices that form part of the NVE are interconnected. Data Filtering subcomponent refers to the filtering of unnecessary (or less important) stored or interchanged information. Data Distribution subcomponent handles the distribution of the data between the NVE nodes. Resource Balancing component manages the assignment of tasks, roles (e.g., Server, Client or Peer) and data management among the nodes. Predictive Modeling and Synchronization subcomponents refer to the time management (i.e., when the data is transmitted and processed, and when the behaviors are executed). Finally, the Computing Models component manages the distribution of the computing and processing load among the nodes. The rationale for dividing an NVE into these components, a more detailed description of them and their relevance in NVEs is explained later.

Up to date, for creating and implementing NVEs, developers and researchers have been designing and implementing a great number of techniques to solve problems that may arise in each of those components. They include methods, technologies, procedures, and other technical solutions that manage the components, deal with the problems in each of them, make the NVE development easier, and also contribute to the improvement of the NVE users' QoE. As the ecosystem of those existing techniques continues to grow, it is hard to study all of them and decide which ones to select when improving an existing NVE or designing a new one [52].

The main goal of this paper is to review, classify and briefly summarize the existing techniques so far, comparing their advantages and disadvantages, as well as grouping them depending on the NVE component they are included in. The motivation of this work is to provide an updated and useful survey and a new classification scheme of the existing techniques for new NVE developers and researchers. Due to the ever-changing state of and the development of new NVE-related technologies, previous review works on the subject missed the most modern technologies. Moreover, new NVE components have been devised in the last decade, needing new techniques that have not been considered in those previous works. In this paper, an updated and more complete list of NVE components, along with a comparison and classification of the techniques aimed to solve problems in each of them is provided. Additionally, some subfields that are less (or not yet) explored have also been identified as potential areas of interest for future research related to NVEs.

As far as the authors know, this is the most complete survey compiling the existing key technologies in each field, regarding the development of NVEs. As a summary, the main aims of this work are the following:

- To review the up-to-date technologies needed for designing NVEs and compare their advantages and disadvantages.
- To provide a novel taxonomy, grouping the technologies for solving the problems found in each component of an NVE.
- To identify the different subfields that are less (or not yet) explored as potential areas of interest for future research.
- To provide NVE developers and researchers with a useful baseline for developing new NVEs or improving the current ones.

With the increase of the number of NVEs and the multiple solutions proposed so far, the presented survey aims to become a useful tool and a starting point not only for future researchers in NVE design but also for those who are novel in this field.

In order to provide a background knowledge about NVEs to the readers, and to help them to better understand the rest of the paper, before presenting the survey of techniques and their classification, some NVE fundamentals and the different aspects and factors that have an influence in the correct performing of an NVE, affecting the end-users' QoE, are briefly presented.

The rest of the paper is structured as follows. In the next subsection, the fundamentals of NVEs are described and the main NVE-related problems are introduced. In Section 2, a compilation of other surveys related to the classification of techniques for solving those problems is presented. In Section 3, the proposed classification is summarized. In the following sections, different network architectures (Section 4), filtering techniques and data distribution models (Section 5), resource balancing techniques (Section 6), prediction and synchronization techniques (Section 7), and computing models (Section 8) for NVEs are described and compared, providing highlights on fields that deserve further exploration. Conclusions and insights into future work are summarized in Section 9. Finally, as many acronyms are used throughout the paper, in Appendix I, a list of all of them explained is provided.

## 1.1. NVE Fundamentals

To help the reader to better understand the contents of the rest of the paper, the main fundamentals of NVEs are explained in this section. Firstly, the features along some useful definitions are provided, and, secondly, the relevant challenges in NVE development are described.

### 1.1.1. Main features of the NVEs

Even though NVEs can vary significantly depending on their purposes, most of them must provide the distributed users with the following primary features:

- A shared sense of space (awareness). The distributed users should have a common feeling of being in the same space [54].
- A shared sense of presence (presence): Users should feel they are part of the VE and are sharing the space with other users (i.e., they should feel they are physically there) [55].
- A shared sense of time (causality/eventuality) [1]. Users should feel they are experiencing the same events at the same moment, with the same duration, and in the same order of occurrence.
- Fidelity: In realistic NVEs, the environment and its contents should be convincing (i.e., as realistic as possible). For example, the virtual objects should have the same visual details as if they were authentic and behave realistically. This feature would not be considered in NVEs with imaginary VEs [1].
- Interactivity. Users should be able to interact with other users (or their representations) and with objects in the virtual world, if possible, a way as much natural as possible. Users not only watch but intervene, actively taking part in the NVE [56].

These features define what is known as immersivity, which refers to the user's feeling of immersion inside the VE. These features affect the users' QoE, and thus, are of vital importance when designing NVEs (the focus of NVE development should be put on the users' requirements). However, due to the unfeasibility of emulating reality at an exact level, NVE developers try to create an illusion of immersion in the environment. Being able to reproduce a feeling of immersion and natural interactions when users are geographically separated is one of the main goals of NVEs. Nevertheless, this goal is especially sensitive to different issues that will be explained later in this section.

Additionally, there are other features of NVEs that are also important for guaranteeing an acceptable level of users' QoE, such as:

- Concurrency: In a concurrent NVE, users should be able to interact within the VE at the same time (i.e., in parallel) without affecting the final result (i.e. the new NVE state) [57].
- Scalability: It is the property of the NVE to handle the growing of the number of users and its workload [57].
- Flexibility, or Robustness: It is the property of adaptation when (internal or external) changes occur in any component of the NVE. The NVE should cope with errors or unexpected changes or events, affecting its performance as less as possible ( e.g., data recovery solutions should be used when some data have been lost) [58].
- Consistency: This property refers to the requirement that any user's action in the NVE must change the affected entities only in the allowed ways and without any conflict with other users' actions. The actions of all users should not create conflicts and, after the actions take place, every user should perceive the same state of the VE (i.e., there is no discrepancy from what different users see). For example, when a user sees a door open, while another user sees it closed is an inconsistent state [59].
- Responsiveness: This property refers to the specific ability of the NVE to complete users' actions within a given time. The NVE should understand and carry out those actions in a timely fashion. The NVE should deliver the new data (e.g., update messages) to all participants as soon as possible. So, users get fluid interactions [60].

Depending on the purpose of the NVE, it should provide a different degree of immersivity and of the other features. Some examples are an NVE for collaborative work used for designing buildings, which should provide a 3D design VE with high fidelity and need high consistency and responsiveness; an NVE for social interaction, which does not need a VE with high fidelity and may focus on offering high responsiveness and scalability; or a MOG, which may need good robustness and scalability to deal with a high number of simultaneous users.

To ensure the above-mentioned features and the goals of the distinct types of NVEs, developers need to apply many computing-based and communication-based solutions to tackle the related technical problems and, thus, guarantee a desired level of users' QoE. Computing-based solutions are the ones that deal with calculating and processing data (e.g., the rendering of the tridimensional world), while communication-based ones deal with the networking aspect (e.g., the transmission of information between nodes). Although the latter also depend on the former (i.e., apply algorithms and process data), the difference is that the latter focus on solving the problems arisen from using networked systems. Hence, computing-based solutions can be used in both VE and NVEs, while communication-based ones only in NVEs.

When several users share the VE, they need concurrency control, which requires the communication-based techniques managing the consistency and responsiveness for not impairing the users' QoE [61].

In some NVEs, there is a trade-off between consistency and responsiveness features [62]. When developers improve one of both features, the other can worsen. So, they should balance between them depending on the NVE needs. In certain cases, they may want to prioritize one over the other depending on the requirements of the NVE. For example, making the interactions sequential, instead of concurrent, to avoid conflicts between the actions of multiple users (high consistency) leads to low responsiveness because users must wait for the actions of other before interacting with the VE. In order to tackle the consistency and responsiveness trade-off problem, the existing consistency maintenance techniques, dealing with the interactions of users in the concurrent environment, may try to balance consistency and responsiveness in separate ways. Based on how the users' actions are managed, the following approaches can be followed: optimistic, pessimistic, predictive and hybrid approaches [58], [63]. Firstly, in an *optimistic* (or aggressive) approach, actions can be executed without previously checking whether they affect consistency. Each user's local copy (i.e., the data and software representing the environment to the user) of the VE does not wait for his/her interactions to be validated and communicated to the other users but executes them and goes on. In the case that the processed actions cause inconsistencies, rollbacks (the reversal of the NVE state to a previous known consistent one) must be applied to recover the lost consistency. This approach is useful for high latency networks, improving responsiveness, but only when interactions have a low chance to produce inconsistencies, as too many rollbacks will produce the opposite effect, worsening the responsiveness. Secondly, the *pessimistic* (or conservative) approach is, basically, the opposite of the optimistic one. With this approach, each user's local copy of the VE must validate every action, ensuring consistency is maintained, and communicate the new state to the rest of users before allowing further interactions. This effectively limits the responsiveness, but ensures consistency without the need of rollbacks, being a good option when low latency networks are used or when responsiveness is not necessary. This approach is best suited for turn-based games (e.g., a chess online game) or NVEs with a low number of users and geographically close to each other. Thirdly, the *predictive* approach comes between the optimistic and pessimistic ones. In it, interactions are predicted (when they will happen or what their effects will be) before they occur. So, the user's local copy of the NVE does not need to stop to check the consistency (as it can be done in advance) while also reducing the number of rollbacks. Only when wrong predictions are made, rollbacks are needed. This approach is very important, as the previous two ones are not highly scalable. Lastly, a *hybrid* approach combines some of the previous approaches, employing them at different intervals in the NVE life cycle, by, for example, switching from one approach to another when network conditions change, or using different approaches depending on the types of entities (e.g., a vehicle that moves fast may require a higher update rate).

### 1.1.2. Problems to tackle in an NVE

In order to provide the NVE with the explained features in the previous subsection, designers have to face many problems that are briefly explained in this one. Since the NVEs encompass so many components, elements, and fields of study, designers should tackle the specific problems arisen in each part of the NVEs separately (e.g., from hardware devices, software applications, database systems, networks, and timing systems, among many others).

Hardware problems include ensuring the involved devices are fast enough to process the information and are optimized to render the VE with the needed speed depending on the type of NVE. When NVEs are designed for (or also for) wireless devices (e.g., smartphones), contrasting with computers, the designers must deal with their reduced energy capacity, storage, and processing power [64], [65]. Additionally, those devices can face high communication latency, limited available network throughput, and the absence of a shared memory or global clock. Software and, in general, algorithms should be programmed to enable the features of the NVE, like processing the interchanged update messages so that the interactions are synchronized, and to provide feedback at the right moment. Additionally, software is also dependent on hardware, and an NVE that was programmed for a specific hardware may not correctly work with another. For example, an NVE application implemented to run on a desktop computer may not be compatible with mobile devices like smartphones or tablets. Moreover, the database systems face the problems of allowing fast, secure, and reliable data access, as well as of ensuring the integrity and consistency of the stored data. To overcome the problems of limited database storage, reduced computing resources of low-end devices, and software compatibility, nodes can take advantage of the performance of other high-end devices by delegating the storage and computing requirements to other nodes (this refers to the Computing Models component of an NVE).

The networking part includes problems related with communication networks and access devices, which, among others, include latency (delay of update messages), jitter (variability of latency), throughput and loss, affecting mainly the interactivity and the update messages transmission rate [66]–[68]. Different connections with heterogeneous throughput or users experiencing different network latencies can lead to consistency and synchronization problems and, thus, to a worsening of the users' QoE. For example, users with higher network delays will probably interact in an unfair way within the VE since some update messages or events may take more time to reach them. To increase the data transmission rate and reduce the delay and the access time to remote databases, developers should optimize the node connections (Network Architecture component of an NVE), reducing bottlenecks in the network; filter unnecessary information out (Data Filtering component of an NVE), which decreases the network usage; distribute de NVE database between nodes (Data Distribution component of an NVE), so NVE data is closer to the clients; and prevent waiting for update messages by predicting future events and states (Predictive Modeling component of an NVE) of the NVE. Even though NVE providers could avoid those network-related problems in other more straightforward ways (e.g., increasing the available network throughput), in most common circumstances, NVE developers may be unable to control the network (e.g., when connections through the Internet are used).

Another key attribute considered when designing NVEs is time. Even though the meaning of time can vary from one NVE application to another, there are two different concepts of time to be considered: *Absolute Time* (a.k.a. Wall Clock Time) and *Virtual Time* (a.k.a. Causal Time or Simulation Time) [58], [69]. The former is based on the concept of a periodic clock, usually synchronized to the Coordinated Universal Time (UTC). The latter is based on a logical, loosely synchronized clock, as a sequence of ordered events, which halts if no new events occur. Time is relevant to consistency and responsiveness maintenance because both depend on it: responsiveness-related problems can cause increased waiting times between interactions, and consistency problems happen when multiple users interact with the same entities during the same time. To help to solve these problems, the transmission of update messages should be synchronized (Synchronization component of an NVE), so events are processed at the same time by the clients and in the right order, and inconsistencies are reduced. Additionally, NVEs can also predict future changes (Predictive Modeling component of an NVE) and organize node connections (Network Architecture component of an NVE) to improve the responsiveness as well.

Furthermore, due to the unpredictability of networked systems, their usage may cause errors (e.g., network throughput can fluctuate, and programs or servers can fail), impairing the QoE, and therefore, the impact of system errors in NVE should be minimized. So, databases can be distributed between nodes (Data Distribution component of an NVE) to prevent data loss when a node fails; and the required tasks or roles of nodes can also be distributed (Resource Balancing component of an NVE) to be able to manage the computing tasks with higher reliability, and errors affect less the system.

Users are another essential element of the NVEs, as they are the ones experiencing the VE. NVEs should support the number of allowed simultaneously connected users they were designed for. If the NVE is not scalable enough, the number of maximum users interacting at the same time is limited, and the system will fail when more users try to access it (e.g., a MOG server crashing due too many requests, making the service unavailable). For this reason, the organization of node connections (Network Architecture component of an NVE) should be paramount, so the system is able to accept the required number of users, keeping a good level of performance. Moreover, by distributing the management of certain tasks and roles (Resource Balancing component of an NVE) between nodes the NVE scalability also improves. For example, when more peers connect, they can control the synchronization of update messages for their closest zones, improving the NVE management as the number of connected peers grow.

It has been previously described how the improved and new computing technologies allow for better quality and performance, which also impacts the rendering of the environment and the communication between users. Furthermore, the faster the communications, the higher interaction capacity of the NVE, which in turn enhances the sense of presence, immersion and the sense of time as well. Besides this, the shared sense of presence highly depends on the degree of immersion provided by the VE, while the shared sense of time highly depends on the synchronization of the interactions. Finally, the interactions are the base of NVEs, and are very important for communicating with the VE and with other users.

Therefore, the different techniques implemented in the development of NVEs should address all the above problems from all the possible perspectives, with the final aim of providing the users with the best experience possible. Table 1 summarizes some few examples of problems to tackle in an NVEs, indicating their cause, possible solutions and the component these solutions rely on.

| Cause | Problem to tackle | Related NVE Components | Solutions |
|---|---|---|---|
| High network latency | Reduced data transmission and access time to remote databases | Network Architecture | Using different network architectures |
| | | Predictive Modeling | Predicting events before being notified |
| | | Data Distribution | Distributing NVE data among nodes |
| Reduced network throughput | Limited amount of data being transmitted simultaneously | Network Architecture | Using different network architectures |
| | | Data Distribution | Distributing NVE data among nodes |
| | | Data Filtering | Filtering unnecessary information |
| Low consistency | Inconsistent states lead to a bad QoE for users | Synchronization | Synchronizing event transmission |
| | | Predictive Modeling | Predicting events before being notified |
| Low responsiveness | Low interactivity as waiting times increase | Network Architecture | Using different network architectures |
| | | Predictive Modeling | Predicting events before being notified |
| Limited robustness | Failures worsen the user's QoE | Data Distribution | Distributing NVE data among nodes |
| | | Resource Balancing | Balancing the resources among nodes |
| Reduced scalability | Limited maximum number of users simultaneously interacting | Network Architecture | Using different network architectures |
| | | Resource Balancing | Balancing the resources among nodes |
| Limited storage and computing | Low-end devices perform worse | Computing Models | Managing the computing processes |

Table 1. Challenges faced in NVE development.

In this paper a compilation and a novel classification of the different techniques proposed in previous works so far for NVEs, according to the identified components of an NVE (third column of Table 1), is provided. In sections 4 to 8, concrete techniques are briefly explained and compared.

# 2. Comparison with Related Surveys

In the past, some few works have made some efforts for classifying the vast set of existing solutions and techniques designed for, or used in, NVEs, as well as to classify the different components that NVEs comprise. In [70] some components of an NVE are identified, but only few related techniques are classified (e.g., network protocols or data distribution models). Four components are described: Communication, Views, Data, and Processes. The Communication component is related to network issues (bandwidth, latency, and reliability) and the geographic distribution of the users. The Views component is related to the user's viewports of the NVE. The Data component is related to the models of data distribution between nodes. The Processes component is related to the execution of the NVE, including the involved servers and clients, computing requirements, and software.

In [58] and [59] the problems of consistency and responsiveness that can happen in NVEs are studied, dividing the mechanisms to solve them into three main components: Time management, Information management and System Architectural management. The Time Management component includes synchronization, time prediction and concurrency control techniques. The Information Management component includes techniques dealing with network latency, and with data filtering and management methods; and finally, the System Architecture Management component includes network and software architecture, communication protocols, and QoS constraints.

In [63] several techniques for maintaining the consistency and responsiveness in NVEs are classified. A main division between the system architecture and consistency maintenance components is proposed. The system architecture component is divided into the network architecture, data distribution, and communication subcomponents. The consistency maintenance component is mostly presented as a catch-all group containing the synchronization, prediction, interactivity approaches, and concurrency mechanisms.

Finally, in [1], the following basic components of NVEs are defined: Graphic Engine and Displays to visualize the environment, Communication and Control Devices, Processing Systems for the computing and transmission of events, and Data Networks for communication and information sharing. Related techniques/solutions are classified in two groups: Architectures, and Technologies and Protocols. While the Architectures one only includes the basic network architectures (Client/Server, P2P and Hybrid), the Technologies and Protocols one includes 3D Technologies (software, programming languages, frameworks, and interfaces) and several Communication Protocols.

These works have shown their classification of techniques, including the benefits and drawbacks, but focused only on the explored use cases in them. They are quite informative and descriptive, but leave other important fields unexplored, such as, e.g., computing models, or prediction techniques, which are included in the taxonomy presented in this paper. Moreover, in this paper, more modern techniques appeared in the last decade (mostly thanks to cloud technologies) are also surveyed and a more updated and complete classification is provided.

In this paper, the components described in the previous works are also considered, together with new ones that have gained special attention in the last years. For example, Cloud Computing has recently become more relevant, allowing for newer techniques (like remote rendering), fact that adds a new component for NVE development and new techniques to study, not considered in previous works. Furthermore, this paper is also centered on the components that are intrinsic to NVEs in contrast to non-networked VEs, so the presented taxonomy is more NVE specific.

Table 2 summarizes the relation between the NVE components defined in this this paper with the taxonomies of the aforementioned works, comparing the number of techniques for each component that each work classifies.

| Taxonomies | Network architectures | Information management | | Resource balancing | Time management | | Computing Models |
|---|---|---|---|---|---|---|---|
| | | Data filtering | Data distribution | | Predictive modeling | Synchronization | |
| Macedonia et al. [70] | X (3) | X (2) | X (3) | - | X (1) | X (0) | - |
| Delaney et al. [58], [59] | X (3) | X (3) | X (0) | X (1) | X (2) | X (7) | - |
| Fleury et al. [63] | X (3) | - | X (3) | - | X (1) | X (4) | - |
| Bouras et al. [1] | X (3) | - | - | X (0) | - | X (0) | - |
| This paper | X (5) | X (5) | X (4) | X (9) | X (4) | X (12) | X (5) |

Table 2. Comparison with previous related works. An 'X' indicates the work takes the component into consideration and the number inside the parentheses indicates the number of techniques classified by that work in that component. The number '0' indicates the paper identifies the NVE component but does not classify any technique for that component.

# 3. Proposed Taxonomy

In this paper, a novel taxonomy, more NVE-based and suited for NVE design than the ones in the works summarized in the previous section, is presented (shown in Figure 3). It includes most of the techniques used to design NVEs, classified into the following components of an NVE: Network, Information Management, Resource Balancing, Time Management and Computing Models. Each of these identified components is clearly separated from the other components, making it easier to classify the existing techniques included in them. In this section, these components are described, while all the involved techniques in each one of them will be briefly explained in the following sections.

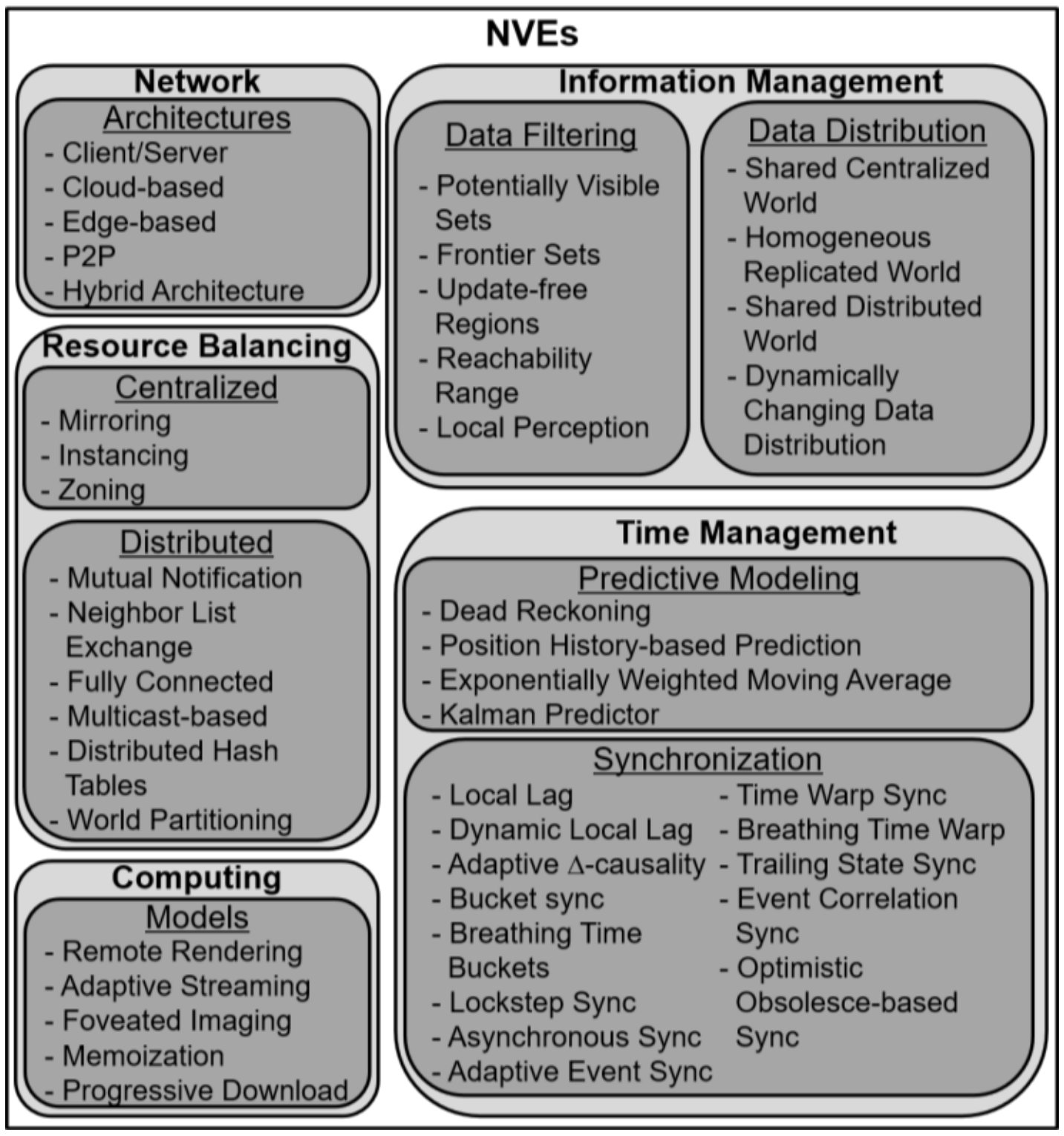

Figure 3. NVE Taxonomy. Components of an NVE and techniques included in them.

With this new taxonomy, a clearer and more intuitive classification than the one of the previous works is provided, making it easier to identify the different components of an NVE, navigate through all the included techniques in them, and quickly find all their relationships. Moreover, the different identified components and subcomponents of an NVE also point out the main research fields related to NVEs (e.g., networking, data distribution models, computing, etc.).

The Network component of NVEs is related to the structure that enables communication between the involved nodes in an NVE, as in [1], [58], [63]. This component consists of two subcomponents: Network Architecture, which defines the connections between nodes; and Communication Protocols, which establish the protocols that the network nodes employ in the communications. In this work, only the network architectures subcomponent is considered, and a classification of the existing solutions so far is provided. As far as authors know, the only existing specific communications protocols designed for NVEs, are DIS and HLA protocols. The DIS protocol was used in NPSNET [24], while the HLA protocol [71] was created to outperform DIS and replace it. The latter is often used for Military applications in private networks. In the rest of NVEs general purpose and widely supported transmission protocols, can be employed, such as TCP, UDP, RTP/RTCP, etc. Due to that, the review and classification of protocols used in NVEs is out of the scope of this paper, and, therefore, the Communication Protocol subcomponent of the Network component of NVEs is not considered.

Regarding the Information Management component of an NVE, it includes all the techniques that directly manage the NVE's data (reading, modifying, copying, or deleting them). Data are one of the more important parts of NVEs [59], [63], [70], [72]. At the same time, this component contains techniques for data filtering and data distribution. Additionally, it can also include the data compression and file systems control techniques [73] as well, but they will not be considered in the classification either, as no solutions specifically designed and applied to NVEs have been found.

The division of the distinct responsibilities that nodes can have in an NVE, to ensure the correct operation of the NVE, is another pillar of NVE development. Different NVEs could identify different resources that must be managed by one or several nodes. For example, one NVE could only require multiple zones to be controlled by the same server and another NVE could allow the same zone to be managed by multiple peers. The techniques that orchestrate the roles between the available nodes are also classified in the Resource Balancing component of the NVE in the presented taxonomy. In previous works, this component is also referred to as Interest Management, Resource Management, or Load Balancing [5], [70], [74]–[76].

The Time Management component of an NVE deals with the execution (or simulation) of the involved processes, with the events and other messages generated and transmitted between nodes, so that actions take place in a causal, coherent, and consistent manner [72], [77]. As NVEs are programs that execute orders over time, and the passage of time is required for users to perceive movement and progress, this component constitutes another pillar of NVE design. Many techniques already exist for time management in NVEs, which can be divided into two groups: Predictive Modeling and Synchronization techniques, which will be explained later.

Finally, the Computing Models component [78], [79], include the techniques that deal with the computation tasks in an NVEs (e.g., rendering a frame), which are important for bringing enhanced functionalities to the distributed users. Previous works on NVEs have barely delved on this subject, and most solutions exist only for MOG applications. In this work, this component is considered as relevant, due to the recent improvements in computation technologies (e.g., faster processors and networks) that go hand in hand with the increase in computation requirements (e.g., more quality and data).

In the following sections, an up-to-date and more complete classification of many techniques for each component of the NVE is provided. Unlike previous works, several comparison tables are included showing their advantages and disadvantages.

# 4. Network architectures

The layout of the network/s involved in NVEs is very important. A well-designed one will perform better by optimizing the network usage and reducing packet transmission load and delay, hence making the consistency easier to control without loss of responsiveness. In this section, the main types of network architectures used in NVEs are described and compared: Client/Server, Cloud-based, Edge-based, Peer-to-Peer (P2P), and Hybrid. Figure 4 shows these architectures.

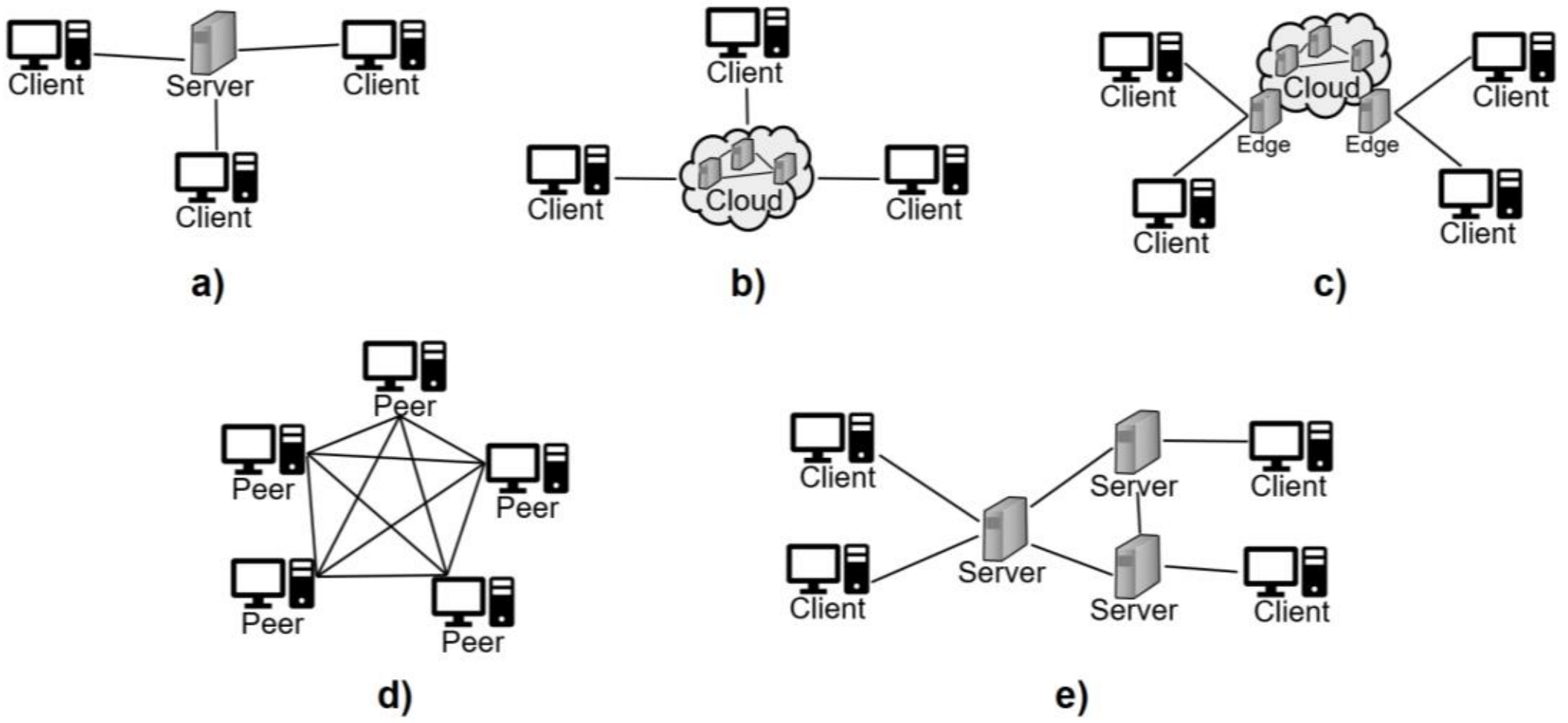

Figure 4. Network architectures: a) Client/Server, b) Cloud-based, c) Edge-based, d) Peer-to-Peer, and d) Hybrid.

## 4.1. Client/Server architecture

In a Client/Server architecture (Figure 4a), a server stores all the virtual environment data and manages the NVE

state and communications between clients. The server acts as a central authority to which all clients must connect and send updates [61]. To communicate or to inform about an event, the clients must communicate it to the server and then it is forwarded to the other clients [59], [63]. Rokkatan [14], RING [26], ShareX3D [29], Co-Surgeon [32], CAVRN [45], TerraNet [80], and the one in [81] are examples of NVEs that follow this architecture.

On the one hand, the main advantages of this architecture are its easy implementation, and a simple consistency, synchronization, and security control, as the server manages the whole NVE. On the other hand, this architecture presents robustness and scalability problems. The server is a weak point, but it can be overcome with the use of redundant or load balanced servers. When the number of clients (i.e., users) increases, a bottleneck can occur on the server or its network access links due to the rising number of communications, and, thus, consistency maintenance can be affected. Finally, it can unnecessarily increase the latency between clients since the communication goes through a server.

## 4.2. Cloud-based architecture

The Cloud-based architecture (Figure 4b) is like Client/Server, but instead of having a single Server, the workload is distributed among several computers with sufficient resources and computing capabilities that manage the needs of the NVE [61], [82]. It is mostly used in massively multiplayer online (MMO) games to solve the scalability problem in the Client/Server architecture. When following Client/Server architectures, game operators are forced to adapt their resources to the game load peaks with the consequent high economic cost and expensive maintenance. Another possibility is to take advantage of the current Cloud Computing business model, which is based on a pay-per-rent model, in which the customers only need to pay for the resources they use. Thus, it is an elastic model that allows NVE operators to adapt dynamically and rapidly the resources used depending on the game load. Some examples of the use of this architecture can be found in WoW [15], in CloudyGame [83], in [84], and in [85].

This architecture presents some advantages, such as the dynamic adjustment of the resources, or the reduction in maintenance costs, and the easiness to maintain consistency. The NVE application must include additional parts to give support to the dynamic provisioning of resources, such as a *monitor* and a *provisioner*. The *monitor* part must consider different performance measures of parameters, such as the response time, the average system throughput, the amount of consumed bandwidth or the use of the rented machines. The *provisioner* part handles analytical load models and fast prediction algorithms to anticipate load peaks and under-utilization of resources. As a drawback, the implementation of this architecture is more complex, since it needs additional parts, and extra delays appear due to the use or access to those parts.

## 4.3. Edge-based architecture

The Edge-based architecture (Figure 4c) extends the Cloud concept, offering the data and the processing resources in Servers closer to the clients (called Edge nodes), improving network usage and responsiveness. In that sense, an Edge node is just a part of the Cloud that interacts with it, doing tasks that benefit the clients close to them. Variations of this architecture can bring to the Edge the parts of the Cloud partially (Fog [86]) or totally (Cloudlet [87]). Examples of the use of this architecture can be found in MUVR [88] and in CloudyGame [83].

Due to the increasing problems of accessing the Cloud with higher amount of data and number of clients, which impair the quality of the service or content, this architecture offers a specialized way of dealing with them and reducing the overhead of the centralized or the Cloud-based infrastructures. The advantages of this architecture are a lower latency, reduced costs, and optimized network bandwidth usage, while its disadvantages are less security and robustness, compared to the Cloud-based one, as the edge nodes are more vulnerable to attacks and failures.

## 4.4. P2P architecture

P2P (Peer-to-Peer) architecture (Figure 4d) is a decentralized architecture with peers interconnected without the need of a central authority or server. The peers supervise and distribute the NVE load between them, having all of them similar roles by acting, at the same time, as clients and servers [1], [61], [63], [89]. Examples of the use of this architecture can be found in MiMaze [11], NPSNET-V [24], DIVE [28], Phaneros [90], SimMud [91], and in Pithos [92].

In general, P2P architectures have some advantages: they provide high scalability (i.e., support a big number of clients) and facilitate local consistency and responsiveness. By contrast, as clients have local copies of the NVE, it is more difficult to keep the global consistency compared to Client/Server architectures. Moreover, due to every client having control of the NVE, security issues may arise (e.g., cheating) [93]. NVEs can tackle this problem by including a node to monitor activity and validate clients' identities.

## 4.5. Hybrid architecture

This kind of architecture mixes other architectures to solve common problems of P2P and Client/Server ones, by combining the centralized and the decentralized schemes (Figure 4e). In hybrid architectures, on the one hand, multiple servers connect themselves using P2P, and, on the other hand, one client connects to only one server in the same way as a traditional Client/Server architecture [63]. To inform clients, servers coordinate themselves. Examples of this architecture can be found in Diamond Park [42], in [94], and in [95].

The main advantages of the hybrid architectures are the ability to provide scalability (as the number of clients increases, more servers can be easily added) and redundancy (i.e., duplication of the NVE data). On the other hand, servers communicate with clients and with other servers, thus, each server must process more data, in addition to the fact that multiple servers can introduce more latency to the NVE.

## 4.6. Comparison

Table 3 presents a summary of the different network architectures employed in NVEs, described in this section, including their advantages and disadvantages, as well as some examples of NVEs following each of them.

| Network Architectures | Advantages | Disadvantages | Examples |
|---|---|---|---|
| Client/Server | • Easy implementation<br>• Easy consistency control<br>• Easy synchronization control<br>• Easy security control | • Robustness problems<br>• Scalability problems<br>• Adds latency | Rokkatan [14]; RING [26]; ShareX3D [29]; Co-Surgeon [32]; CAVRN [45]; TerraNet [80]; Pandzic et al. [81] |
| Cloud | • Easy consistency control<br>• Easy robustness control<br>• High scalability<br>• Reduced maintenance cost | • Increased latency<br>• Complex implementation | WoW [15]; CloudyGame [83]; Nguyen et al. [84]; Nae et al. [85] |
| Edge | • Reduced latency<br>• Reduced network usage<br>• Reduced maintenance cost | • Global consistency problems<br>• Less secure than the Cloud one | MUVR [88]; CloudyGame [83] |
| P2P | • Easy local consistency control<br>• Easy local responsiveness control<br>• High scalability | • Global consistency problems<br>• Less secure than centralized | MiMaze [11]; NPSNET-V [24]; DIVE [28]; Phaneros [90]; SimMud [91]; Pithos [92] |
| Hybrid | • High robustness<br>• High scalability | • Adds data usage and latency<br>• Complex implementation | Diamond Park [42]; Anthes et al. [94]; Capece et al. [95] |

Table 3. Network architectures used in NVEs.

The way the nodes interconnect is highly related to the techniques that are needed to guarantee and maintain consistency in the NVE. For example, when using a centralized architecture like Client/Server, the techniques that require peers are discouraged, while a Hybrid or P2P one will take advantage of those techniques. Additionally, networked systems may also experience scalability, flexibility, or security issues, which should be balanced or managed by those architectures (e.g., a Client/Server implementing an authentication system on the server to improve security), or by other techniques explained in the following sections.

## 4.7. Future research directions

Network architectures can be centralized, distributed, or a combination of both and, in this sense, the field of network architectures is already explored. Nonetheless, with the advent of ubiquitous computing, thanks to Internet of Things (IoT) and Edge computing, new network architectures can be devised and applied specifically to NVEs [96]. Therefore, newly created network architectures can be expected soon.

# 5. Information management

One of the most important parts of NVEs is their data, how they are stored and how they are transmitted between the partakers. In this section, Data Filtering and Data Distribution techniques for optimizing the data of the NVE and for reducing the network usage in it, respectively, are presented. Data filtering techniques are the ones that select the needed data to be transmitted or processed, while data distribution techniques are the ones that manage the replication of the database of the NVE (or specific parts of it) among the participant nodes.

## 5.1. Data filtering techniques

Filtering data at transmission, reception or in any of the intermediary nodes can reduce traffic overload and increase scalability [97]. To do so, data filtering techniques can be used to set a priority for the updates of the generated states and, if needed, to discard the transmission of data that is deemed as less important. This way, when a lot of events happen and not all the data could be processed or transferred, the network overload does not increase. However, the extra processing load required for filtering data increases latency and, hence, causes transmission delays that reduce responsiveness. So, the goal of the data filtering techniques is to reduce network congestion and usage that could cause large delays, at the expense of some temporal inconsistencies, given that those inconsistencies are barely perceived by the users. In this subsection, some data filtering techniques are explained and compared, such as: Potentially Visible Sets, Frontier Sets, Update-free Regions, Reachability Range and Local Perception.

### 5.1.1. Potentially Visible Sets (PVS)

PVS is built on the basis that an avatar can only see a set of the total entities in the virtual world. So, only those ones in that set should be rendered [98]. The virtual world is divided into zones and, for each one, all the entities that can be viewed from any point of that zone are stored. Static entities can be used and stored beforehand, whereas dynamic ones (e.g., which can change their position) must be re-processed when they change. Clients, instead of being constantly (every frame) calculating visibility, only need to receive updates of the set of entities visible from the zone in which their avatars are located. An example of the use of PVS can be found in Phaneros [90], and in [99]. Figure 5 shows an example of a world divided into 9 zones. When the avatar moves from one zone (a) to another zone (b), its visible entities change.

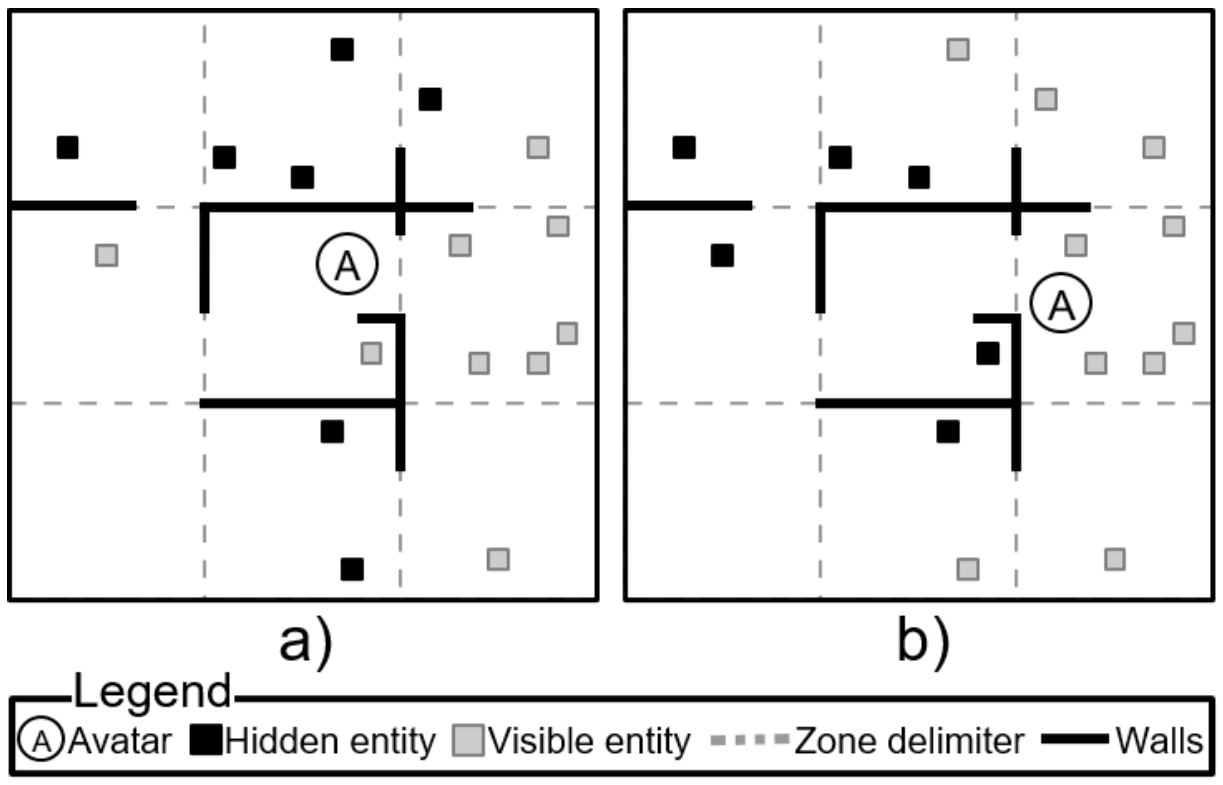

Figure 5. Virtual world divided into 9 zones with PVS. The avatar moves from a) to b), changing the zone and, thus, the visible entities.

PVS is simple to apply and can save data processing and network usage when there are few visible entities. Nevertheless, it requires extra storage and, in open virtual spaces or with a lot of dynamic entities displayed at the same time the data size increase can become uncontrollable.

### 5.1.2. Frontier Sets

Frontier Sets is based on PVS. The NVE is also divided into zones, but, in this case, the clients look for other visible zones instead of visible entities, so that all the entities inside a visible zone are considered, even if they are hidden. Frontier Sets consists in finding, for each zone containing an avatar, a pair of groups of zones in which one group has no visibility in common with the other [100]. A group of zones is called a Set, and a pair of sets, in which an avatar located inside one set cannot see the contents or entities located in the other set, is called a Frontier. Ultimately, all existing frontiers are called the Frontier Set of the NVE. When an avatar enters a zone in any set, the frontier sets including that set are calculated and, the corresponding client only receives update messages from the zones visible from that set. This way, an avatar can move freely within a set, without the need of requesting information from other non-visible sets. An example of the use of Frontier Sets can be found in [101], and in [102]. As an example, Figure 6 shows a couple of frontiers. An avatar in the zone 4 could combine frontiers a) and b) to ignore the zones 3, 5, and 9. If the avatar moves to the zone 7, it will leave frontier a), being able to view the zone 9.

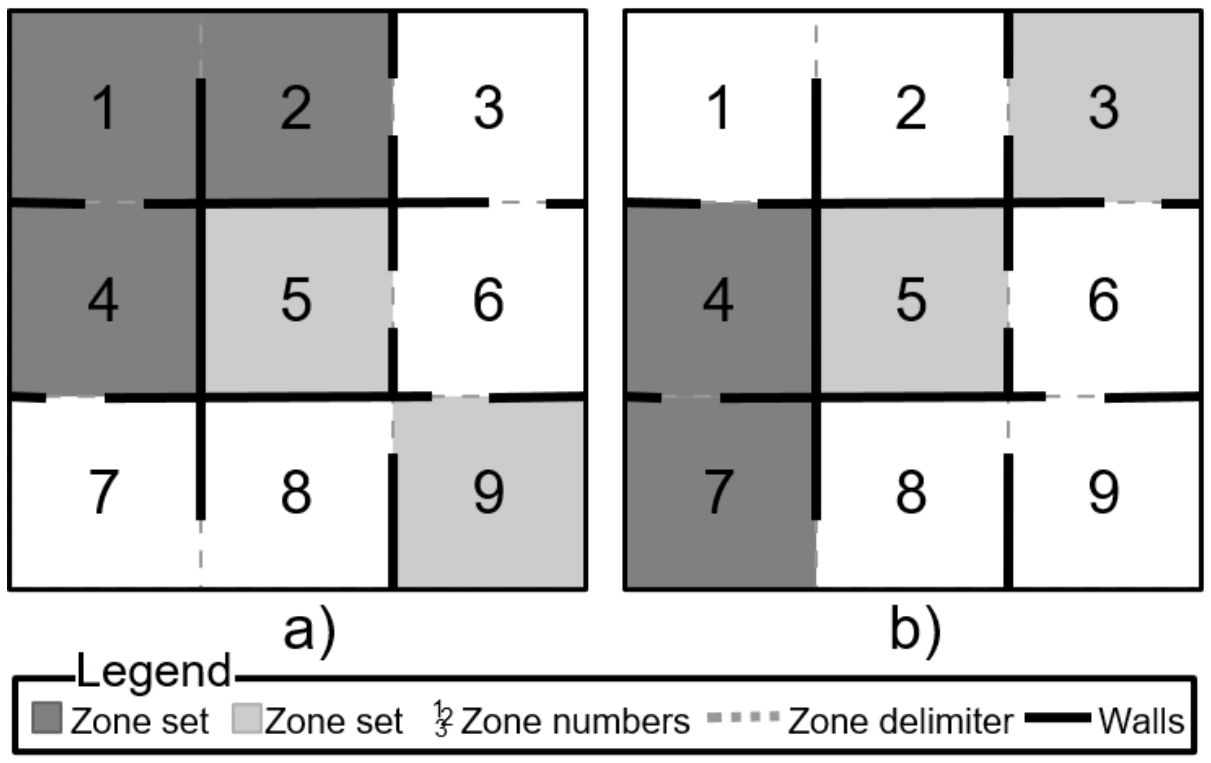

Figure 6. Frontier Sets. In a), the sets of 1-2-4 and 5-9 zones. In b), the sets of 3-5 and 4-7 zones.

In Frontier Sets, each client checks all the frontiers of the zone in which its avatar is located to know from which other zones the user does not need to receive updates, instead of calculating the visibility for each entity (as in PVS). Thanks to that, the required memory and processing resources of the NVE do not increase when the number of entities increases. Nevertheless, if the number of entities is high inside one zone, the number of messages transmitted could still be high and the zone should be further divided (increasing the storage need) or the consistency should be ensured by other means.

### 5.1.3. Update-free regions (UFR)

In UFR, zones are called regions, and the NVE defines, for every possible pair of entities, a pair of regions, so that the contents of one region are hidden for the other region in the pair [103]. Both regions are considered update-free regions (UFR) in the pair since clients with avatars in one region will not receive updates from the other region in the pair. While an avatar stays inside a UFR in a pair, the associated client will not send update messages to the clients with avatars in the other region of the pair. An example of the use of UFR can be found in [103].

Although UFR may be less robust than frontier sets, it does not require knowledge of the whole contents of the NVE by each client, as they only keep track of the zones they have to leave, instead of all the possible zones, or entities in sight [59]. This makes UFR suited for distributed architectures like P2P.

Figure 7 shows an NVE with two avatars, each one in a different UFR (separated by a wall) in the same pair of regions. The clients of each avatar will not receive updates from the other while their avatars remain inside their respective regions. When the avatars leave their UFR, they check if they became visible to each other, and a new pair of UFRs will be generated for that pair of avatars so when the avatars enter and stay inside them, they will not send update messages to each other.

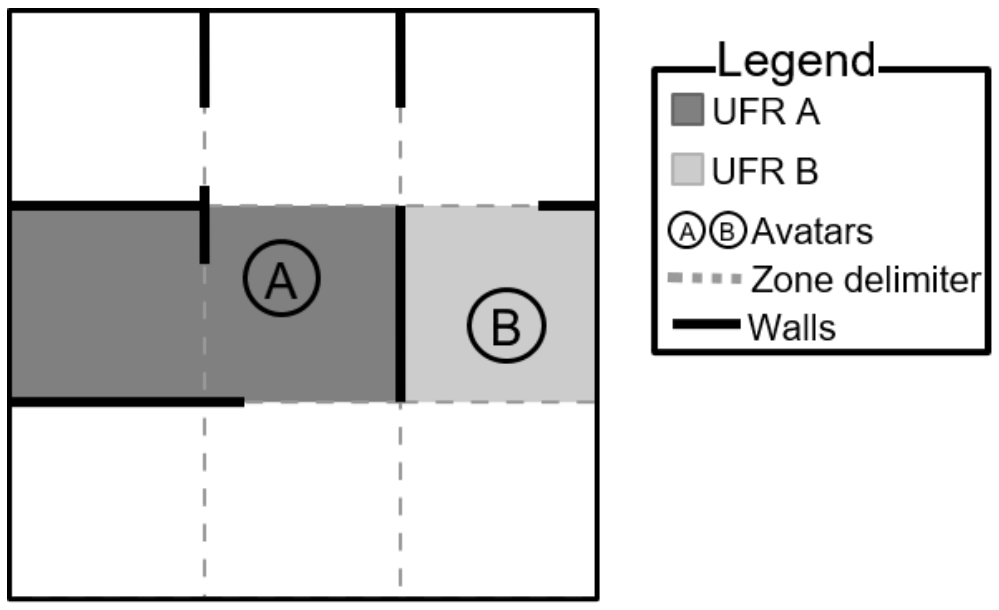

Figure 7. Example of Update-Free Regions.

The main benefit of UFR is that the network usage is reduced by not sending unnecessary updates, and without affecting the consistency perceived by the user. The UFRs are easy to compute, but as the number of entities increase, the size of memory needed to store the data of the generated UFRs increases, becoming hard to be managed for all the possible pairs of entities.

### 5.1.4. Reachability Range

In Reachability Range, instead of dividing the virtual world into zones, a circular zone is defined around each avatar and the associated client only accepts update states from entities inside that zone (i.e., only entities inside are updated) [97], as shown in the example in Figure 8. An example of the use of Reachability Range can be found in Second Life [40], Pithos [92] and in [97].

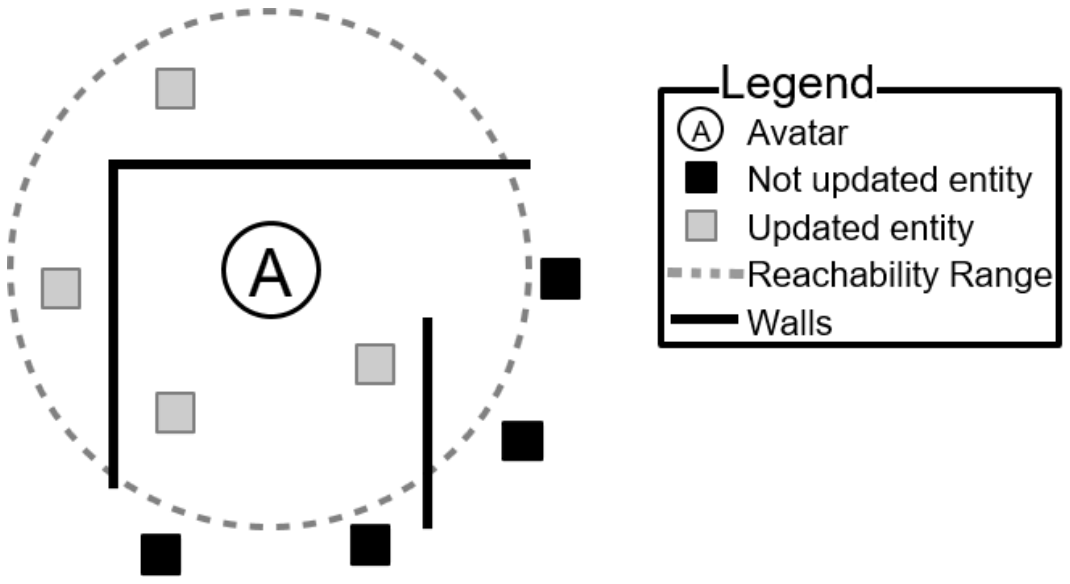

Figure 8. Example of reachability range (nearby entities are still updated even if not visible).

Reachability Range is easy to implement, but its performance depends on the number of entities within the defined range, as too many could slow down the NVE or reduce the consistency. In this case, no computation is required to determine pairs or zones, saving memory as well.

### 5.1.5. Local Perception

In Local Perception, messages are prioritized based on how close or far the entities are in the virtual world from the avatar controlled by a client [104]. All the state changes will arrive to the client but ordered depending on how close the entities are from the avatar. The farther the entity, the later the update will be received. Local Perception is like the Reachability Range, but instead of discarding information of entities beyond a range, that information will be received with delay, and therefore, the corresponding updates of those entities will be delayed. Local Perception effectively distorts time, as the closest entities are quickly updated, while the update of the ones far away is delayed. This can be perceived as a bad effect (e.g., as entities move away from the avatar, the movement slows down), but it allows to receive events from more and distant entities in comparison to Reachability range. This way, if an entity is getting closer, its state will be updated more frequently, so no jumpy movements are perceived. Figure 9 shows, from a) to c), how an entity that is approaching the avatar is perceived by the client, and its real position. The closer the entity, the sooner it is updated, closing the gap between the real position and the received one. An example of the use of Local Perception can be found in [104].

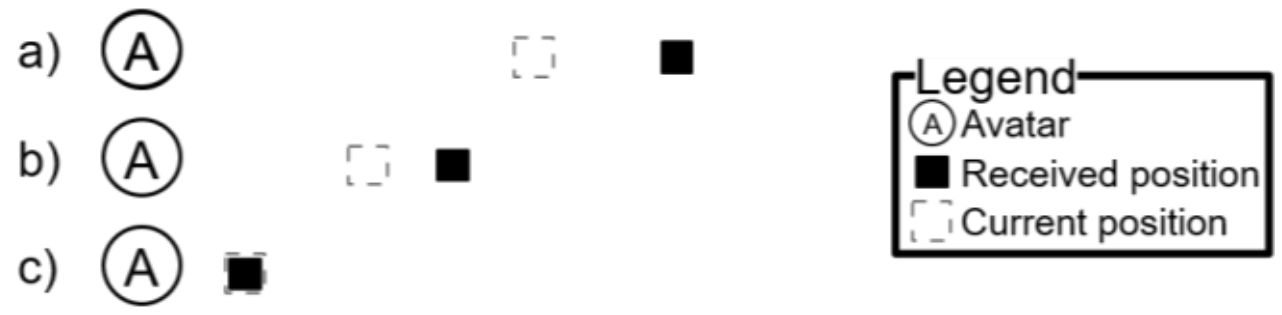

Figure 9. Movement of an entity (dashed square) and the state received at the time (black square). The closer to the avatar, the more accurate its position is.

As an advantage, Local Perception supports a high number of entity updates without a high loss of consistency and responsiveness. On the contrary, when there are a lot of entities, update messages can be delayed more than expected, even causing network congestion.

### 5.1.6. Comparison

Table 4 presents a summary of the different data filtering techniques described in this section, including their advantages and disadvantages, as well as examples of NVEs using each of them.

| Data Filtering | Advantages | Disadvantages | Examples |
|---|---|---|---|
| Potentially visible sets | • Easy implementation<br>• Reduced network usage | • Increased storage need<br>• Scalability problems | Phaneros [90]; Moreira et al. [99] |
| Frontier sets | • Reduced storage need<br>• Scalable for dispersed entities | • Scalability problems for high number of entities | Avni et al. [101]; Steed et al. [102] |
| Update-free regions | • Reduced traffic overload<br>• Reduced computation needs | • Increased storage need | Makbily et al. [103] |
| Reachability range | • Easy implementation<br>• Reduced computation needs<br>• Reduced storage need | • Scalability problems for high number of entities | Second Life [40]; Bassiouni et al. [97] |
| Local perception | • High scalability<br>• Delayed consistency | • High traffic overload<br>• High network latency | Sharkey et al. [104] |

Table 4. Data filtering techniques used in NVEs.

### 5.1.7. Future research directions

Notice that the found techniques are focused on data filtering based on position or visibility. Filtering techniques yet to be developed could be dynamic or based on predictions. As the data filtering techniques specify the condition from which to filter in or out the update messages, dynamic filtering techniques could adapt these filtering conditions depending on the requirements and the current state of the network. For example, if the network throughput is reduced, more update messages should be filtered out, and when it is recovered, less messages may be filtered out. Moreover, prediction techniques could also determine whether certain update messages are more or less important than the rest, so they could be prioritized and the impact on consistency would be minimized.

## 5.2. Data distribution techniques

Choosing the location of the data is a critical decision when designing an NVE. Data are usually collected in a DB containing all the information about the elements composing the virtual world, the position of the avatars, the model geometry, textures, terrain, and their behaviors (i.e., the way they react to an event). The locations where the data are stored and how they are replicated among the participants can reduce network usage and network latency. In this subsection, several methods to distribute and replicate the data among the NVE nodes are explained and compared, such as: Shared Centralized World, Homogeneous Replicated World, Shared Distributed World and Dynamically Changing Data Distribution.

### 5.2.1. Shared Centralized World

In Shared Centralized World, a server stores the DB and shares the contents with the clients (Figure 10, with the different shapes inside the DB standing for different entities). Clients must connect to the server to be able to interact within the NVE [70]. Every time the state of an object is going to be modified, a request must be sent to the server. Then, the server performs the changes in the DB and sends an update message to all the other clients to update the state of that object (Figure 10). Shared Centralized World is frequently used in Client/Server architectures, and some examples can be found in [105] and VISTEL [106].

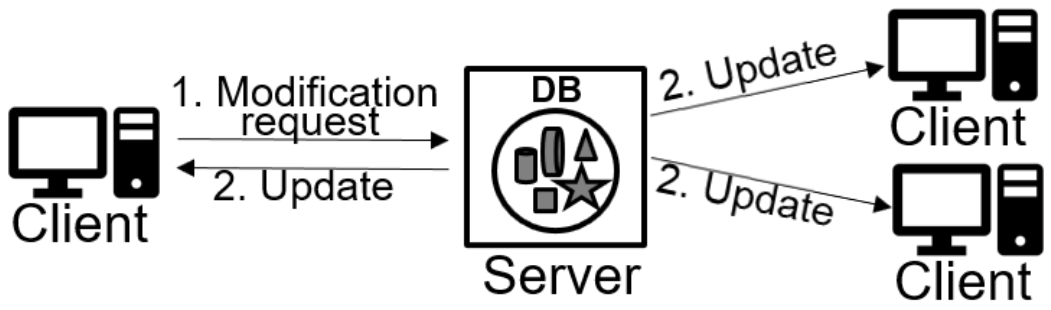

Figure 10. Shared Centralized World, with the steps of a modification.

The main advantages of Shared Centralized World are the ensured consistency, as only one database is used, and the absence of data replication on several servers. On the other hand, in addition to the well-known drawbacks that are present when using a Client/Server architecture (e.g., robustness, scalability…) this mode presents two additional main drawbacks. First, possible high transmission delays between clients and the server, and the processing time in the server can increase the latency, inducing a lack of responsiveness, worsening the interaction and, therefore, becoming annoying for users. Second, a bottleneck might occur in the server, as the higher the number of clients, the higher the number of messages to be processed in it and the higher the number of update messages to be sent by it.

## 5.2.2. Homogeneous Replicated World

In Homogeneous Replicated World, each client stores a copy of the virtual world DB (which is modified locally) and has the control of the object behaviors [1], [70] (Figure 11). Only object state changes or events (e.g., collisions) are interchanged between clients, to maintain consistency. When a client changes an object behavior (e.g., opening a door or cutting down a tree), synchronization techniques must replicate these behaviors in the rest of the clients (Figure 11). Examples of Homogeneous Replicated World can be found in SIMNET [16], RING [26] and ShareX3D [29].

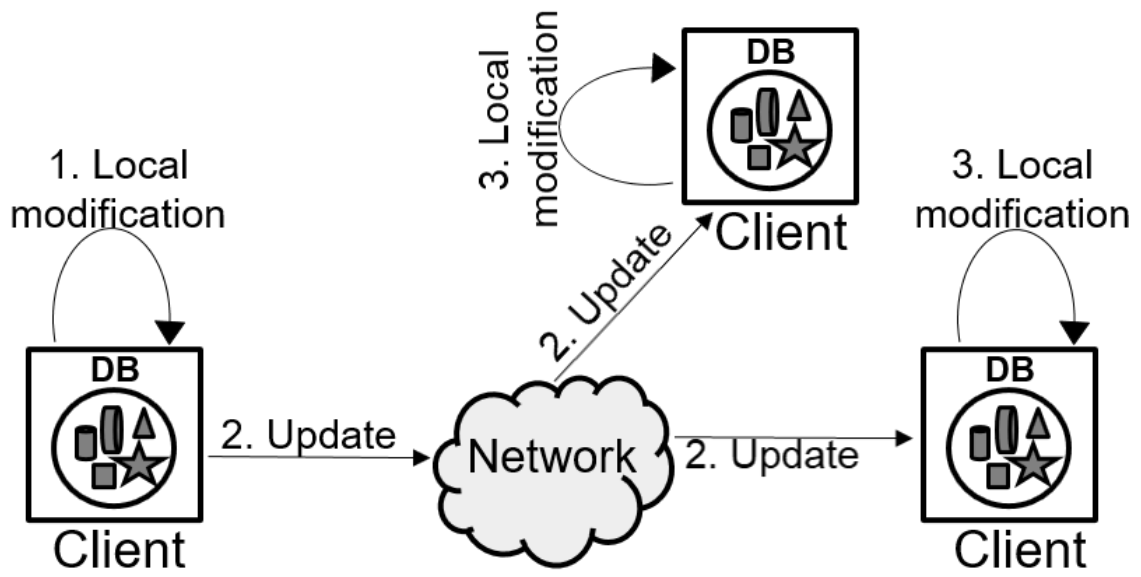

Figure 11. Homogeneous Replicated World. All the clients store the same DB; and every time an event is triggered, each DB is updated and synchronized.

The use of Homogeneous Replicated World presents two main benefits: the sent messages are simple updates, and their size and number are quite small; and the latency on the interactions is very low. Furthermore, any modification to the objects in the virtual world is performed by clients. On the contrary, it has some few drawbacks. First, some inconsistencies can occur between the clients since message losses or network delays could prevent some clients from updating their own copy of the DB on time. Moreover, additional mechanisms are needed on each client to manage the concurrent access. Each user can locally modify the environment but only when this modification is transmitted, possible conflicts can be detected. Another flaw is the size of the DB, since the bigger the NVE the bigger the amount of data to be stored. Finally, the lack of flexibility is another important disadvantage. Adding new elements to the NVE can be a hard task as they must be created and replicated into all the DB copies in each client.

## 5.2.3. Shared Distributed World

Unlike in Homogeneous Replicated World, in Shared Distributed World, the clients do not store a copy of the full database, but each one stores only a different part of it that is shared with the rest of the clients [70] (Figure 12). So, these clients act as the servers of their DB. Examples can be found in DIVE [28], RAVE [107], and BrickNet [108].

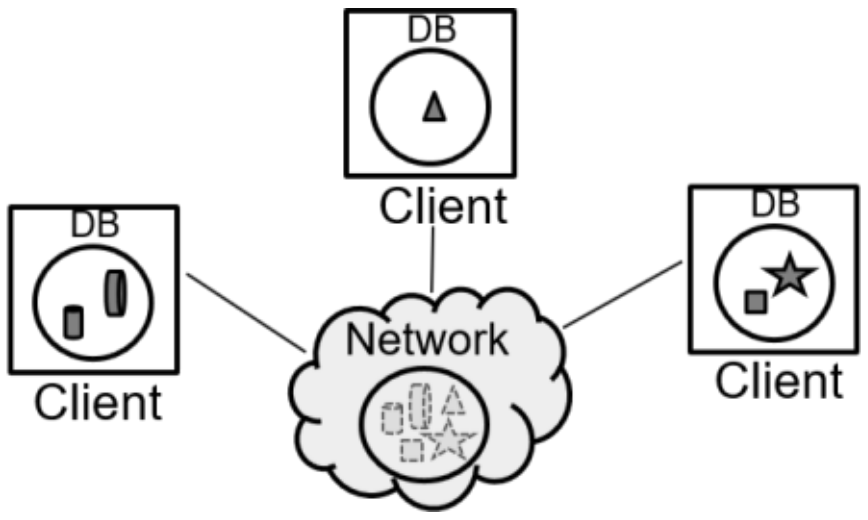

Figure 12. Shared distributed world (clients storing different parts of the DB and sharing them through the network).

By using it, the data filtering process and the initial database download can be reduced, saving time and network usage. Needed resources are more reduced and keeping consistency in the NVE becomes an easier task than in Homogeneous Replicated World. The consistency in the complete database, however, needs to be guaranteed by using other methods (e.g., by using Dynamically Changing Data Distribution, presented in the next subsection, or a hybrid network architecture).

### 5.2.4. Dynamically Changing Data Distribution

This is a hybrid data distribution technique, which dynamically adapts the replication of data to be either Shared Centralized or Shared Distributed, depending on the requirements of consistency and responsiveness of the entities. In the same NVE, some entity manipulations require good responsiveness (e.g., real-time movement) while others require strong consistency (e.g., turn-based event) [72]. Data distribution can also be changed depending on the capabilities of the network or its nodes. This way, depending on the object the user is interacting with and the current network latency, the data distribution can be dynamically changed to make for a trade-off between consistency and responsiveness. An example of the use of this data distribution technique can be found in COLLAVIZ [34] and in OpenMask [109]. In it, clients have a replicated copy of the NVE data. Nonetheless, each of the object behaviors will be executed in just one of the clients, denominated controller of that object. In other words, each object is assigned to only one controller. Two types of handles are used: reference object handle and mirror object handle. Reference object handles are used by the controllers to identify simulated objects whose behaviors they can execute or control (these objects are named referents). Mirror object handles are used by the controllers to identify additional copies of their referent objects but controlled by other clients (i.e., copies of the objects whose behaviors will be executed in a different controller). Figure 13 depicts the process of altering an object. If the referent object to be modified is in the user's client, that object will be locally manipulated and then an update message will be sent to the rest of the clients which have and control a copy of that object (steps 1 and 2 of Figure 13). On the other hand, if the object to be modified by the user is not a referent object in the user's client, a request will be sent to the remote client, which is the controller for that (mirror) object (steps 3 to 6 of Figure 13). In that remote client, the object will be modified, and an update message will be sent to all the other clients controlling copies of that object.

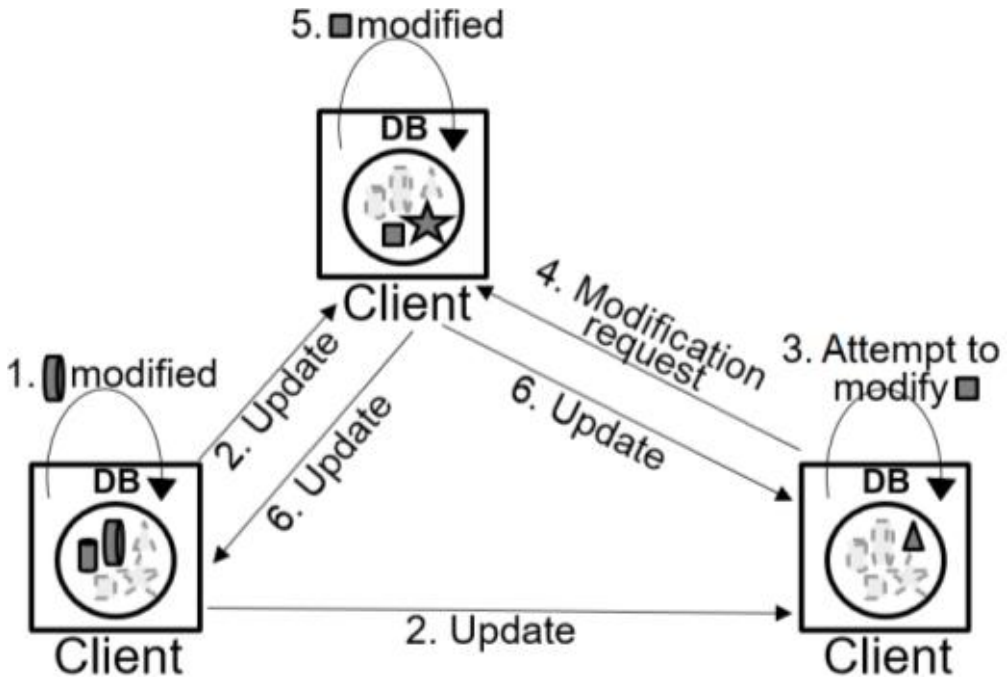

Figure 13. Dynamically Changing Data Distribution. A client modifies a local entity, while another modifies a mirrored entity.

The advantages Dynamically Changing Data Distribution offers are a better scalability compared to centralized solutions, and that the NVE can minimize the throughput requirements. On the other side, this technique is more complex to implement, and the interchange of additional control messages can affect the network throughput.

### 5.2.5. Comparison

Table 5 presents a summary of the data distribution techniques presented in this section, including their advantages, disadvantages, as well as examples of NVEs using them.

| Data Distribution | Advantages | Disadvantages | Examples |
|---|---|---|---|
| Shared centralized world | • Easy consistency control<br>• Reduced storage need | • Responsiveness problems<br>• Scalability problems | Curtis et al. [105]; VISTEL [106] |
| Homogeneous replicated world | • Easy responsiveness control<br>• Reduced network usage<br>• Reduced network latency | • Consistency problems<br>• Concurrency problems<br>• Robustness problems<br>• Increased storage need | SIMNET [16]; RING [26]; ShareX3D [29] |
| Shared distributed world | • Easy local consistency control<br>• Easy responsiveness control<br>• Reduced storage need | • Global consistency problems | DIVE [28]; RAVE [107]; BrickNet [108] |
| Dynamically changing data distribution | • High scalability<br>• Reduced network usage | • Complex implementation<br>• Adds control messages | COLLAVIZ [34]; OpenMask [109] |

Table 5. Data Distribution in NVEs.

### 5.2.6. Future research directions

The explained data distribution techniques cover all the possibilities for the storage and replication of the NVE data, from the most centralized database to the most distributed ones, including hybrid solutions as well. Evidently, centralized data distribution techniques are more suited for centralized network architectures, and distributed data distribution techniques are better used with distributed network architectures. Furthermore, and like the network architectures, it is important that future research on the field of data distribution in NVEs focuses more on the specific characteristics unique to NVEs (e.g., considering the entities and real-time interactions). For example, instead of just replicating arbitrary binary data of databases, specific implementations could consider a group of data belonging to specific entities, avatars, zones, and so on, when each could be managed independently from the others (a user's client requires the data of the zone, where his/her avatar is, to be available).

# 6. Resource balancing

To address the scalability problems of certain techniques, the NVEs can distribute their computation load and, hence, the tasks needed for running the NVE applications, among several nodes, by using resource balancing techniques, depending on the architecture of the NVE [5], [110]. This is quite different from the information management techniques, which have direct control to the data and not to the computing requirements. The resource balancing techniques contribute to reduce the network usage and the end-to-end delay, making the NVE more scalable, but may add problems related to security, robustness, or global consistency maintenance. Moreover, they can be centralized, when controlled individually (by a server or a sole peer), or distributed, when several peers manage the same processes.

## 6.1. Centralized balancing

Centralized techniques for resource balancing are mainly suited for centralized architectures (Client/Server or Cloud-based ones) [5]. They grant authority to the same node or group of nodes (e.g., multiple servers sharing workload and different tasks between them) for managing a session for a group of clients that share the same interests (e.g., the players of the same match, a group of clients with avatars in the same zone, or all the attendees of a virtual event). Different centralized resource balancing techniques are described and compared in this subsection, such as: Mirroring, Instancing and Zoning.

### 6.1.1. Mirroring

This consists in dividing the management of overloaded NVE zones into a set of servers. Each one of these servers manages and processes the computing, states, and a part of the DB for different subgroups of entities placed inside the overloaded zone (Figure 14). The information and updates of all the entities in that zone are then synchronized among the servers. Examples of the use of Mirroring can be found in [14], [111]. Figure 14 shows two servers, each managing different entities (represented with different shapes inside the DBs) for the same NVE that a client accesses by connecting to each of the servers, while the servers connect between them for interchanging control messages.

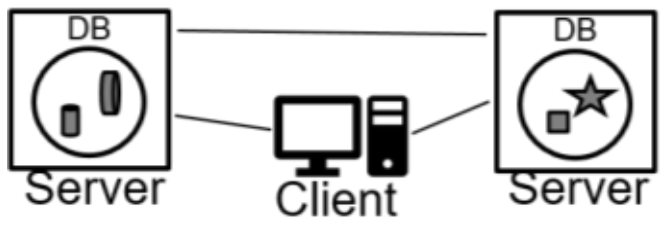

Figure 14. Mirroring. Two servers, each managing different parts of the NVE (represented with the different shapes) for the same overloaded zone.

Mirroring efficiently manages the network usage, adapting it to the overload of zones in the virtual world. Nevertheless, it adds extra computing and design complexity requirements with the need of synchronization when the number of users increases.

### 6.1.2. Instancing

It is a simplification of Mirroring, in which session load is distributed by starting multiple independent instances of the same NVE zone [112]. These instances are independent of each other, thus, players in different instances cannot interact and do not see each other even if their avatars are closely placed in the virtual world (Figure 15). Thanks to this, the NVE is easily scalable. However, this solution cannot be used when it is needed that the users view and interact with all the other users in the same virtual zone at the same instant.

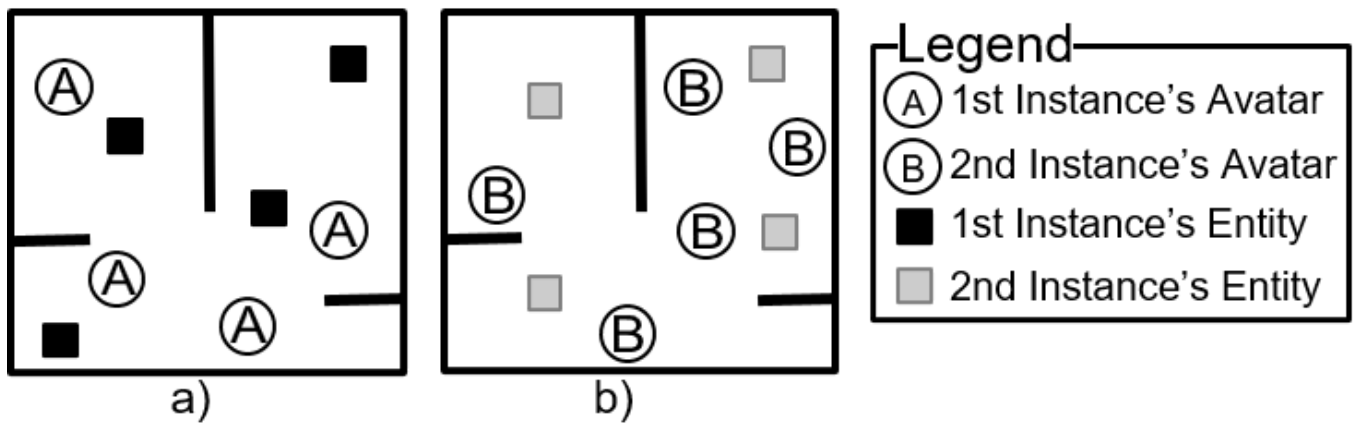

Figure 15. Instancing. Two instances of the same zone, with different entities.

Instancing is mostly applied in MMO games, such as, e.g., CoH [12] and WoW [15]. In WoW, it is referred to as sharding. Different instances of a zone are generated and the players within each zone are distributed between instances depending on different features of the game, such as their level, progress, or party (grouped in-game players). In that game, a technique called cross-region zoning is also employed to reduce the number of instances, when needed. If the number of users inside some instances decreases, multiple instances can be merged into one, balancing the number of players in them. This way, the number of users in each instance is always balanced.

### 6.1.3. Zoning

In Zoning, the virtual world is divided into different zones that are handled independently by separated servers [75], [113]. Clients then connect to the server that holds the zone their avatars are located and interact between them through the same server. When an avatar moves to another zone, the server changes (Figure 16). An example of the use of Zoning can be found in [113], [114], and [115].

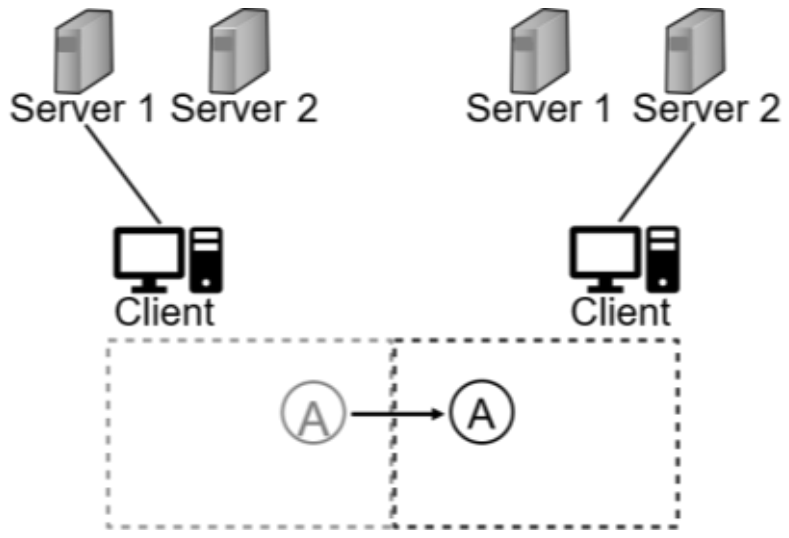

Figure 16. Zoning. The client changes the server when its avatar moves from the left zone (controlled by Server 1) to the right one (controlled by Server 2).

The overload of the NVE is distributed, making it better manageable and easier to implement. Nevertheless, controlling the zones can become a difficult task when the number of clients fluctuates a lot.

## 6.2. Distributed balancing

Distributed resource balancing consists in changing the manager of the processes and resources as avatars enter, move through, or leave the different zones of the NVE or the near surroundings of other avatars. These techniques are based on distributed network architectures, like P2P, and, therefore, cannot be applied in NVEs following centralized network architectures [104], [116]. The techniques that are explained and compared in this subsection are: Mutual Notification, Neighbor List Exchange, Fully Connected, Multicast, Distributed Hash Tables, and World Partitioning.

### 6.2.1. Mutual Notification

With Mutual Notification, clients are directly connected only to those other clients (peers) whose avatars are closely located in the virtual world (considered as neighbors). Only when a peer's avatar changes its position, interact with the virtual world, or a new peer's avatar enters in the surroundings of another peer's avatar, notifications between direct neighbors will be sent, so they can achieve global connectivity and neighbors' discovery. Thanks to that, network usage is optimized, but at the expense of adding the overload of calculating the neighborhood. Additionally, having a lot of close users in the virtual world can impair the consistency or responsiveness. In Figure 17, it can be seen how only peers with their avatars close enough in the same zone of the virtual world get connected (3 groups of peers).

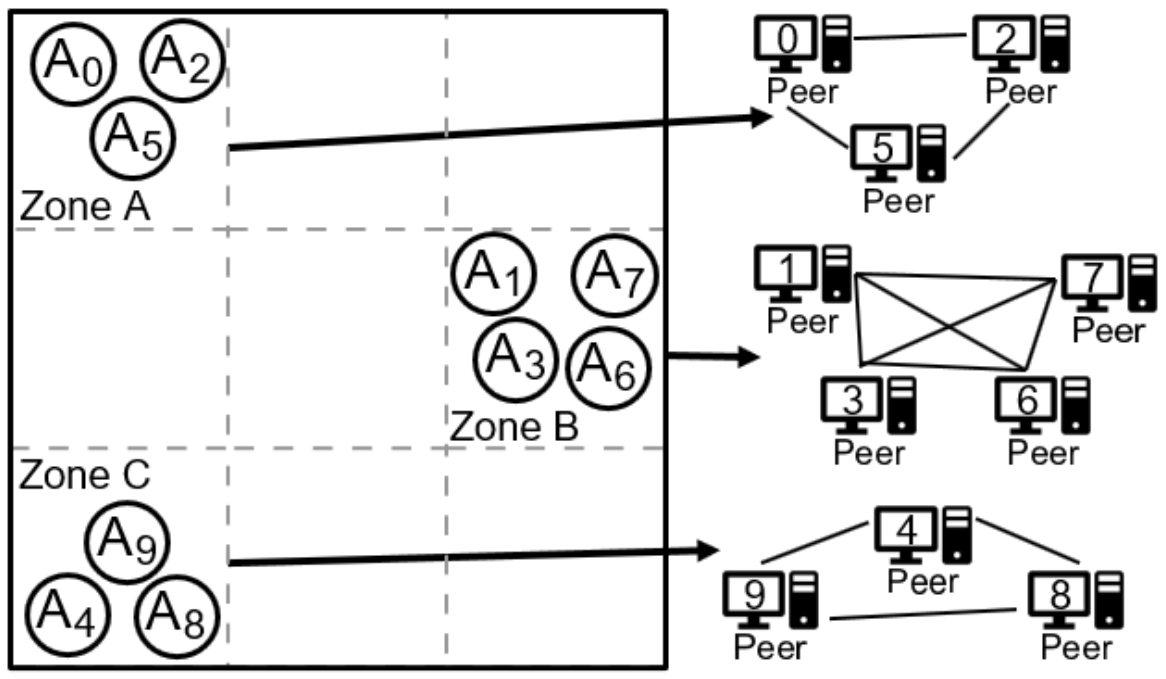

Figure 17. Mutual Notification in a virtual world (10 peers) with 3 groups.

Examples of the use of Mutual Notification can be found in VON [117] and in pSense [118]

### 6.2.2. Neighbor List Exchange

When using Neighbor List Exchange, peers keep knowledge about the existence of their neighbor peers' avatars (in the virtual world) and about the avatars of the neighbors of the neighbors by interchanging messages continuously (Figure 18). To get a better view of the virtual world and to be better informed of the state changes and updates, a peer receives all the information from its neighbor peers [119]. These neighbor peers not only send information about their own avatar and its actions (e.g., updated states on entities modified by its own avatar) but also retransmit the information received from their own neighbor peers (e.g., updated states on entities modified by its neighbor peers' avatars).

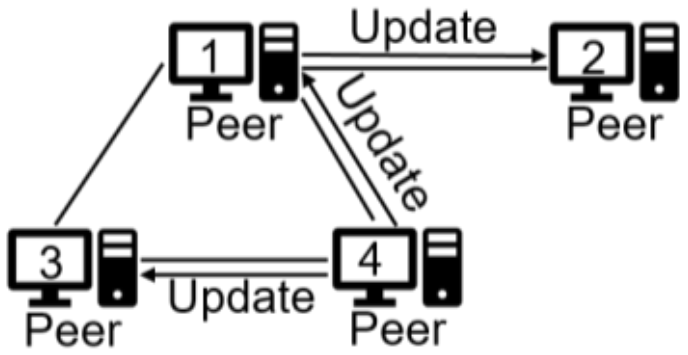

Figure 18. Neighbor List Exchange with a peer sending an update message.

Figure 18 shows the peer number 4 sending update information to the rest of its neighbor peers. Peer number 1 retransmit the update information to its own neighbor, the peer number 2. Examples of NVEs can be found in [120] and [121].

Neighbor List Exchange can reduce the network usage, but the interchange of messages can affect the NVE negatively, and the global consistency cannot be guaranteed.

### 6.2.3. Fully Connected

This consists in connecting every peer to each one of the others, so they interchange the updates directly. This way, delays between the participants are reduced. This has been done in some systems intended to improve the performance of MOGs, such as Pithos [92], and Donnybrook [122].

When the number of connected peers increases, scalability problems arise. With the purpose of having a more efficient NVE, data filtering techniques can be applied, limiting message transmissions.

### 6.2.4. Multicast-based technique

This takes advantages from IP multicast delivery, as well as application layer multicast (ALM). A multicast group is created with peer clients that share an interest in common entities or zones of the virtual world. Only members of that group will receive updates about state changes of those entities or zones. Some examples of NVEs employing Multicast are NPSNET-V [24], DIVE [28], TerraNet [80], and SimMud [91].

Multicast scales well when the number of clients increases but can experience some flexibility and fairness issues when peers connect through heterogeneous access networks, with different bitrates and latencies.

### 6.2.5. Distributed Hash Tables (DHT)

This is an effective and fair way of balancing the system load among all the peers. In general, the entities in the virtual world are divided into groups and each group is assigned to one peer participating in the NVE. Each peer has an ID and manages a group of entities. To decide the group of entities each peer manages, a hash function (to associate different variables with a certain length value) is used (Figure 19). This process has two steps. Firstly, a hash function is used to convert data of variable length to fixed size data. The parameters used by the hash function are usually the latency and the geographical location of the peer. Secondly, values produced by the hash function that are similar are associated to the same peer. For instance, in Figure 19 it is shown how closer positions and latencies produce similar values. SimMud [91] and Colyseus [123] are NVE systems which use DHT but combined with multicast. Another example can be found in [116].

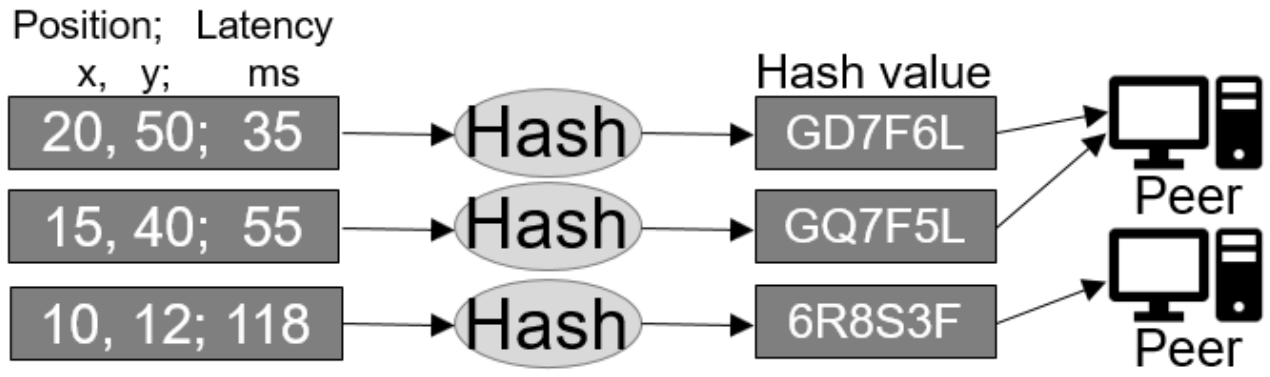

Figure 19. Distributed Hash Tables. Position and latency are passed through a hash function and entities are assigned to peers based on the returned value.

DHT improves the NVE scalability and robustness and reduces network delays. Nevertheless, as it can be expected, it is more complex to implement than other simpler methods.

### 6.2.6. World Partitioning

The virtual world is divided into zones, each assigned to a specific peer whose avatar is inside the zone, with sufficient computing power, throughput, and storage capacity, known as Super Peers (SP) [124], [125] (Figure 20).

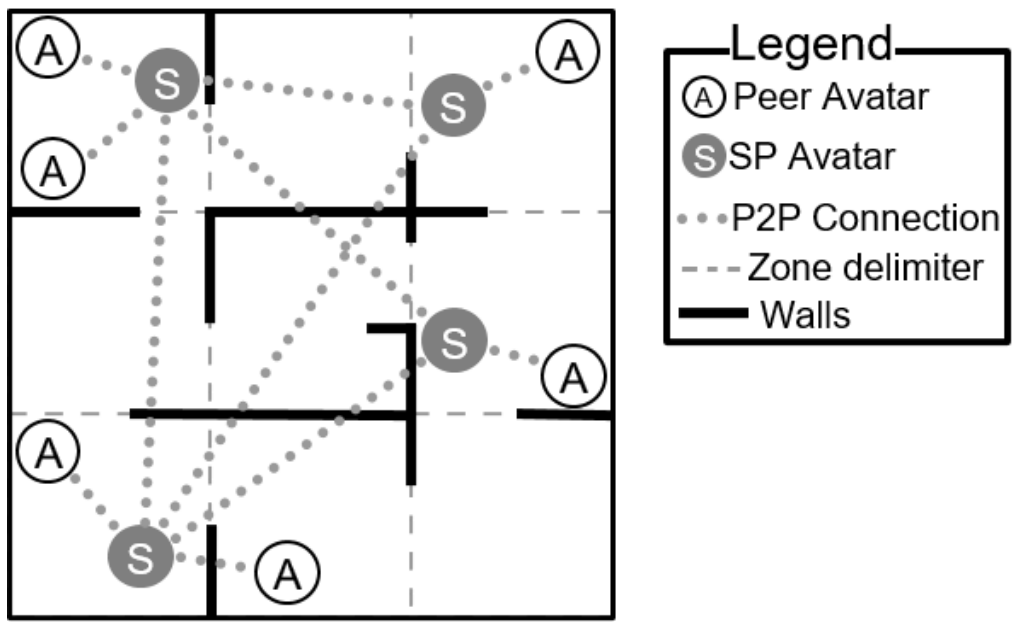

Figure 20. World Partitioning (zones and avatars assigned to an SP in each).

The SPs act like servers and clients at the same time, controlling the zone management (i.e., data from entities in that zone are managed by the SP). However, an avatar's information can be managed by both the SP and the associated peer to which the avatar belongs. Examples of NVEs with World Partitioning are found in Pithos [92], MOPAR [126], and Badumna [127].

World Partitioning is like mutual notification in that both divide and group peers, based on the zone their avatars are located, but, while with mutual notification the peers of the avatars in the same zone get connected one to each other, with world partitioning the peers get connected to the same SP. This reduces computing needs for the group management, compared to mutual notification, as the neighborhood calculation processes a lower number of connections with the peers. Additionally, it also offers a good scalability, and the consistency is easy to maintain. Nevertheless, it is not suited when clients with low capabilities are mostly used (e.g., mobile phones), as it makes it harder to manage the NVE by few peers.

## 6.3. Comparison

Table 6 shows a summary of the explained centralized and distributed techniques for resource balancing used in decentralized NVEs, their advantages and disadvantages, as well as examples of NVEs using them.

| **Technique** | | **Advantages** | **Disadvantages** | **Examples** |
|---|---|---|---|---|
| **Centralized** | Mirroring | • Reduced network usage | • Synchronization problems<br>• High computing needs | Rokkatan [14]; Morillo et al. [111] |
| | Instancing | • High scalability | • Limited number of users | CoH [12]; WoW [15] |
| | Zoning | • Easy implementation | • Scalability problems | Abdulazeez et al. [113]; Cai et al. [114]; Zhang et al. [115] |
| **Distributed** | Mutual notification | • Reduced network usage | • Consistency problems<br>• Responsiveness problems | VON [117]; pSense [118] |
| | Neighbor list exchange | • Reduced traffic overload | • Consistency problems | Chen et al. [120]; Kawahara et al. [121] |
| | Fully connected | • High responsiveness | • Scalability problems | Pithos [92]; Donnybrook [122] |
| | Multicast | • High scalability | • Robustness problems<br>• Fairness problems | NPSNET [24]; DIVE [28]; TerraNet [80]; SimMud [91] |
| | DHT | • High robustness<br>• High scalability<br>• Reduced network delay | • Complex implementation | SimMud [91]; Colyseus [123]; Wang et al. [116] |
| | World partitioning | • Easy consistency control<br>• High scalability | • Robustness problems<br>• Bad when there are only lightweight clients | Pithos [92]; MOPAR [126]; Badumna [127] |

Table 6. Resource Balancing in NVEs.

## 6.4. Future research directions

Resource Balancing techniques can be either centralized or distributed, being suited for centralized or for distributed network architectures, respectively. In general, these techniques already deal with all the kinds of applications and network architectures the NVEs can have, but, like data filtering, newer techniques could be based on more parameters than the position of avatars and node geographic locations. For instance, avatar mobility patterns (i.e., how users move their avatars inside the NVE) could be considered for deciding how the nodes will connect and when the connections should change [74], [128]. For instance, the size of the zones and the number of close entities may affect the way the avatars move through zones, and hence, the avatars could be stopping on certain spots, and moving faster on others. In conclusion, this field could be further explored by studying the distribution of entities inside the VE so new techniques take that information into account to design new techniques.

# 7. Time management

NVE systems need to allow users in a concurrent environment to perceive and interact within the virtual world despite them experiencing different network delays. In this sense, time management solutions deal with the time instants or periods when messages are transmitted and/or processed, to balance between consistency and responsiveness, and include Predictive Modeling and Synchronization techniques. Predictive modeling techniques employ information from previous instants and manage the time in a predictive manner (e.g., predicting future events), while synchronization techniques work in a more deterministic way and are used for ensuring the causality or consistency of the events in the NVE.

## 7.1. Predictive modeling

Predictive modeling techniques try to predict the behaviors of users and entities, and their consequences, to reduce the need to send update messages or to optimize their delivery. Consequently, the traffic overload and network usage are minimized, and end-to-end latency is decreased, while keeping a certain degree of tolerable consistency. The prediction of events can happen in the client originating the event or in the rest of clients that are supposed to receive the associated update message. In this subsection, the following predictive modeling techniques are explained and compared: Dead Reckoning, Position History-based Prediction, Exponentially Weighted Moving Average and Kalman Predictor.

### 7.1.1. Dead Reckoning

In Dead Reckoning, a client that is modifying the state of an entity (e.g., moving an avatar) also runs a simulation, in background, of what the rest of users are perceiving based on the last update message communicated to them until the user modifying a state notifies a new change of the state to the rest of the users [59]. For example, when an entity is moving in a direction at a certain speed according to the last update, it is supposed that it will keep that speed and direction until a change is notified. If the actual state of the entity (e.g., speed, direction, acceleration, position, etc.) exceeds a certain allowed threshold, or degree, of inconsistency (divergence), compared to the simulation (e.g., the current velocity of an avatar is much lower than before -i.e., when the last update message was sent-), a message is sent to update it, and then the rest of users will change the state to the actual one, and the client modifying the state of that entity will go on simulating what the others perceive, from that last update message.

With Dead Reckoning, a margin of consistency is ensured, but the complexity of the NVE is increased. An example of the use of Dead Reckoning can be found in MiMaze [11], SIMNET [16], TerraNet [80], and in [129].

Figure 21 shows Client 2 simulating the movement of Client 1’s avatar, from the last known direction. Meanwhile, Client 1 is moving the avatar and running a simulation, at the same time, of the movement that Client 2 should be viewing. At some point in time (t1, in Figure 21), Client 1 changes the direction of the avatar, so the movement is not the same as the simulation of what the Client 2 perceives. Nonetheless, Client 1 does not send any update message until the actual movement and the simulated movement diverge, exceeding a threshold. This way, when using the Dead Reckoning, a tolerable difference between any state of an entity and its simulation reduces the need for updates.

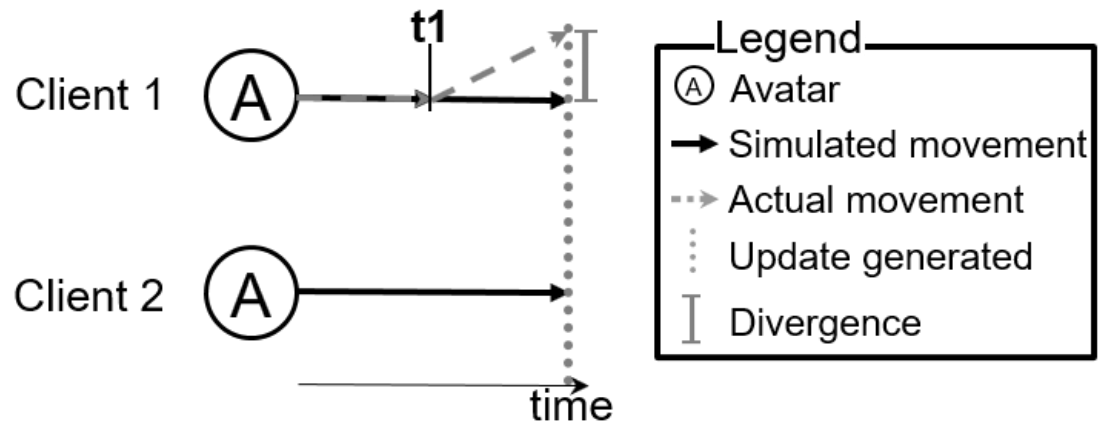

Figure 21. Dead Reckoning (predicted and real movements of an avatar with the diverged distance).

The main benefit of Dead Reckoning is that update messages are only sent when the consistency diverges, exceeding a threshold, and thus, network usage is reduced when participants can estimate the states with a tolerable degree of consistency. Nevertheless, if the states fluctuate a lot or are unpredictable (e.g., avatars making arbitrary movements or entities with high speed), messages will be interchanged more often, and other techniques will be required to handle the inconsistencies and reduce the number of updates needed.

### 7.1.2. Position History-based Prediction

Position History-based Prediction is used to anticipate the movement of entities in other clients in a way that it seems close to the real movement experienced in the client that originated that movement, even with network delays. To do this, the other clients extrapolate future positions using the previous directions and, when an update is received with the correct position, instead of instantly moving to that position, the movement is changed so the entity converges to that position. An example of the use of Position History-based Prediction can be found in [130]. Figure 22 shows an example where an update message for a moving entity is received, the received and the previous known directions are used to estimate the current direction (e.g., an average direction) it is following until receiving the next update.

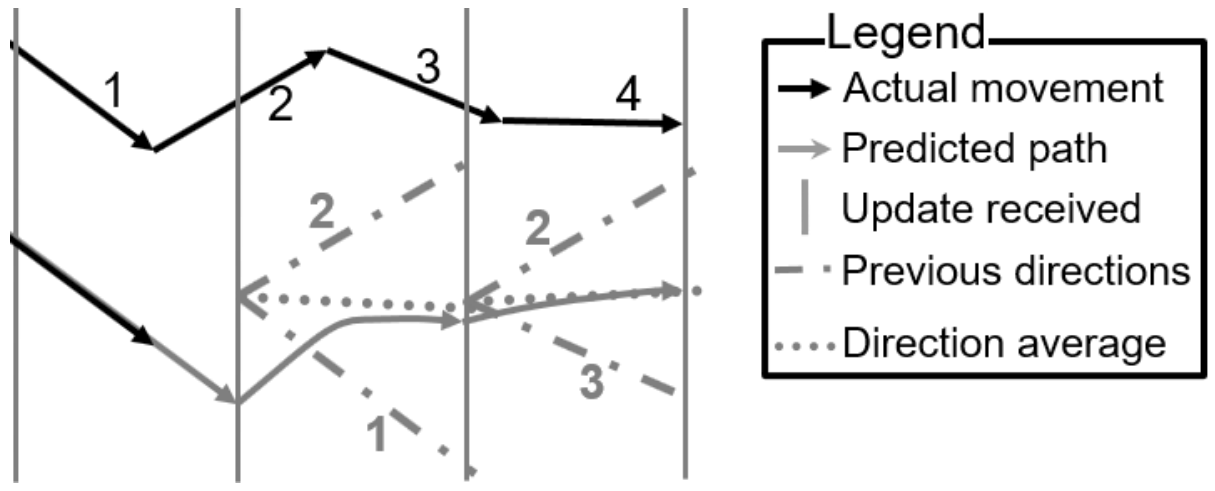

Figure 22. Position history-based prediction (from two previous directions).

With Position History-based Prediction, users will still see a continuous movement, even when no updates are received. Nonetheless, the memory and processing requirements of the NVE are increased and, as it uses past movements to predict new ones, it may lose effectiveness when arbitrary movements happen.

### 7.1.3. Exponentially Weighted Moving Average (EWMA)

EWMA is inspired by the previous one, but it sets different weights to the previous directions, so that the more recent ones are weighted higher [131]. Movements are a combination of position, direction, and speed. The last known movements are used to estimate future movements, favoring the recent ones in the prediction. An example of this predictive model can be found in [132]. Figure 23 shows an example comparing Position History-based Prediction in a), and EWMA in b), in which the movement of an entity is approximated to a direction based on weighted previous directions, so the most recent ones have greater influence.

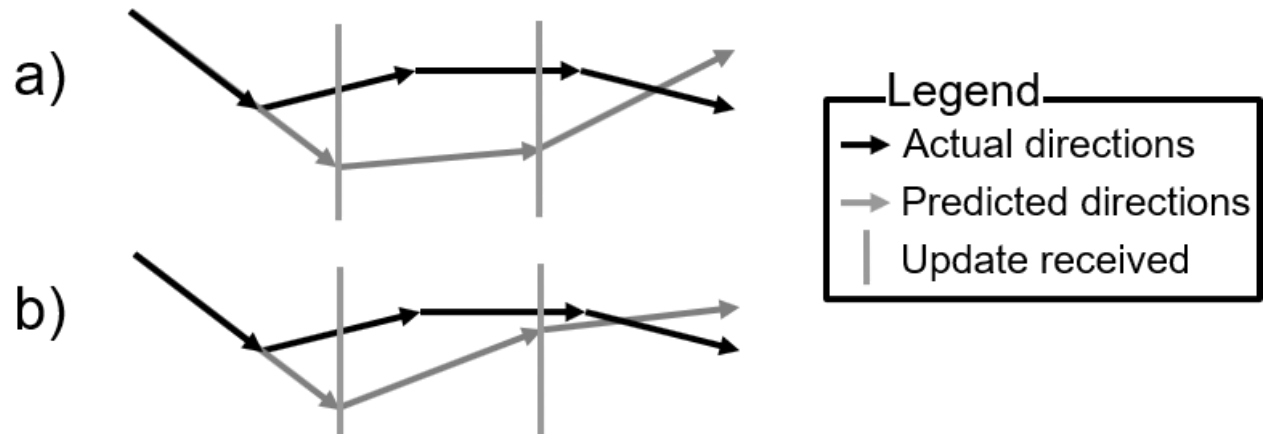

Figure 23. Comparison of Position History-based Prediction (a) and EWMA (b).

EWMA provides a quick adaptation to the specific way entities move (e.g., if an entity has been moving in circles recently, EWMA predicts it will keep moving in that circular course), reducing the inconsistency without increasing the network usage. Nonetheless, as Position History-based Prediction and EWMA use past movements to predict new ones, they may lose effectiveness when arbitrary movements happen. Also, they increase the memory and computing requirements, used to store previous directions, and calculate the predicted ones.

### 7.1.4. Kalman Predictor

The Kalman Predictor is used for reducing the inconsistencies of the tracking prediction between the original entity movement and the simulated one in the rest of the clients [132]. Kalman Predictor has two phases: in the first one, it makes a prediction based on the previous states, while in the second phase, when receiving the actual state, it sets a weight on that prediction depending on how accurate it was (the more accurate, the higher weight). Then, the process is repeated, using the previous weighted predictions for the next estimations. This is represented in Figure 24. In a), the Kalman Predictor is not employed, and therefore, the system does not learn from previous inaccurate predictions. In b), a previous prediction with high error is processed to set new weights, and a latter prediction is weighted according to that, obtaining a more precise prediction.

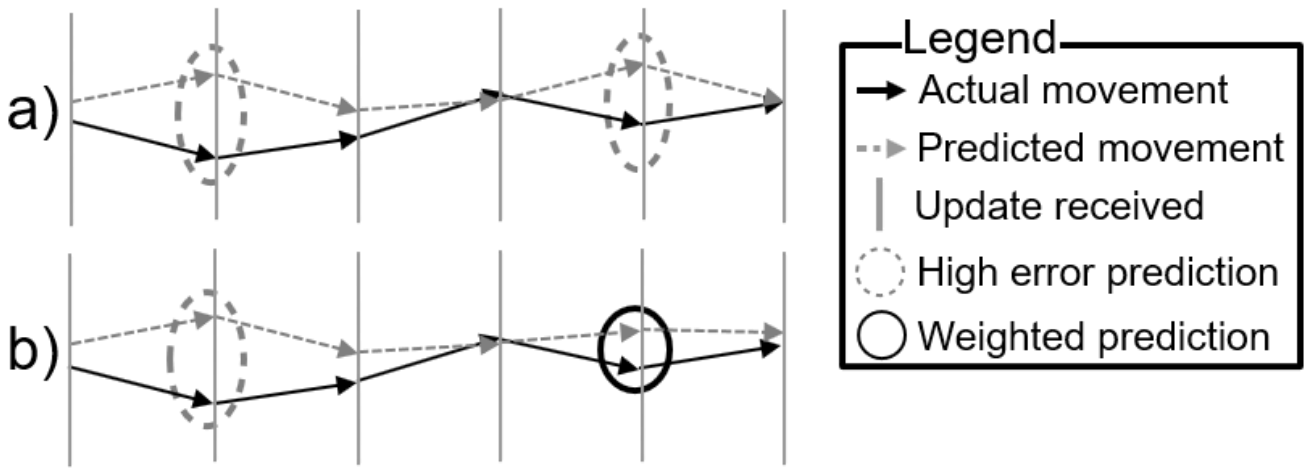

Figure 24. Kalman Predictor improving predictions from previous errors.

The Kalman Predictor adapts easily, requiring less update messages and correcting errors. Like Dead Reckoning, Kalman Predictor is also useful when the movements of the entities are stable (i.e., with low variations). Nonetheless, when movements are fast and arbitrary, Kalman Predictor loses accuracy.

An example of the use of this predictive model can be found in [133], where rather than predicting current positions on the destination client, it is used for predicting future positions at the time the update will be received due to network delays. For instance, if there is a 100ms delay, the sender will estimate the position at 100ms latter and transmit it.

Another example can be found in [134], where it is used to estimate the head motion of users' HMDs.

### 7.1.5. Comparison

Table 7 presents a summary of the predictive modeling techniques used in NVEs described in this section, including their advantages and disadvantages, as well as some examples of NVEs following each of them.

| Technique | Advantages | Disadvantages | Examples |
|---|---|---|---|
| Dead Reckoning | • Reduces the network usage | • Cannot handle high divergences | MiMaze [11]; SIMNET [16]; TerraNet [80]; Chen et al. [129] |
| Position History-based Prediction | • Reduces the network usage<br>• Can predict big deviations | • Cannot estimate arbitrary actions<br>• Increased computation needs | Singhal et al. [130] |
| EWMA | • Reduces the network usage<br>• Adapts quickly to changes | • Cannot estimate arbitrary actions<br>• Increased computation needs | Chan et al. [132] |
| Kalman Predictor | • Adapts quickly to changes<br>• Reduces the network usage | • Loses accuracy on arbitrary actions<br>• Increased computation needs | Tumanov et al. [133]; Gül et al. [134] |

Table 7. Predictive modeling techniques in NVEs.

### 7.1.6. Future research directions

On the one hand, Predictive Modeling techniques deal with the consistency in the NVE, and, as it can be expected, their consistency management approach (conservative, aggressive, etc.) enters in the group of prediction-based approaches. On the other hand, these techniques predict only movement of entities, but newer techniques (e.g., using Artificial Intelligence-based approaches), yet to be developed and applied in the field of NVEs, could be used to estimate other parameters like entity access (i.e., if an entity is going to be interacted with by several clients), the user density of zones (i.e., when a zone is going to be crowded, considering other techniques, like distribution), future network conditions (so other techniques could be adapted), and other interactions like when an avatar will be shooting a gun or opening a door.

## 7.2. Synchronization

In NVEs, the synchronization (abbreviated as sync, hereinafter) techniques constitute important mechanisms to maintain a satisfactory level of consistency and fairness, which contributes to provide truly engaging and interactive experiences to users, despite the existence of network issues. These techniques schedule the notification and execution of events to be performed on specific times. Additionally, they may offer either a *delayed global consistency*, when users perceive the same consistent world but at different times, or an *imposed global consistency*, when the execution of events happens for all users at the same instant. Moreover, sync techniques for NVE can be employed to either synchronize the occurrence of events, when their temporal relationship should be maintained between clients, or to synchronize media streams (composed of stream media units or MUs), which can be continuous streams (e.g., audio or video streams) or data streams containing parts or the totality of one or more events.

Well-known sync techniques in media communications usually deal with the sync of the playout of media streams (i.e., stream MUs inside data packets) [135], [136]. These sync techniques, at the same time, are divided into four types [56], [137]: intra-stream sync, inter-stream sync, inter-destination media sync (IDMS), and inter-device sync (IDES) techniques.

Intra-stream sync handles and maintains the temporal relationship within each time-dependent media stream (i.e., received stream MUs, are processed and presented in the correct order and timing). Inter-stream sync handles and maintains the sync between the playout processes of related (time-dependent or not) media streams (e.g., audio-to-video sync, or lip-sync), i.e., those streams can be played out on the same device or on different devices, which, in turn, can be either close-by (a.k.a. IDES) or far apart, in different locations (a.k.a. IDMS or group sync).

On the one hand, in NVEs, IDES techniques handle the sync between different devices used in one client, such as HMDs, haptic devices, smartphones, smart TVs, and computers, to maintain interactivity and a good QoE when multi-device scenarios are required. Group of devices synchronization techniques (IDES or IDMS) can be divided into three groups, according to the sync control scheme followed by the sync solution [138]–[140]: Master-Slave, Synchronization Maestro, and Distributed Control schemes.

Master-Slave scheme (M/S), which consists of selecting one device as the master device while the other devices are considered as slave ones, and only the master device sends timing information about its playout processes to the slave ones that take it as the sync timing reference.

Synchronization Maestro Scheme (SMS), in which there is a sync maestro device, which can even be an independent device, in charge of collecting timing playout information from all the involved devices, processing it, and sending messages with a calculated sync timing reference to all of them to make them adjust their playout processes to be in sync.

Distributed Control Scheme (DCS), in which all the devices interchange their timing playout information and individually calculate asynchronies between them and adjust their own playout process to be in sync.

In a previous authors' work in [138], a qualitative comparison of the three schemes used for IDMS is presented.

On the other hand, IDMS is the most important sync type in NVE deployment as it is related to more parts of the NVE, which affect the consistency and responsiveness between the partakers of the NVE. Such parts are the end-to-end network, clients, servers (if they exist), and NVE data. Furthermore, this type of techniques are basically the ones in charge of synchronizing the virtual world and the states of its entities between several distributed clients. So, apart from the above group sync techniques (M/S, SMS and DCS), in this section other group sync techniques used in the past for NVEs are considered and compared, such as Local Lag, Dynamic Local Lag, Adaptive Δ-causality, Bucket Synchronization and Breathing Time Buckets, Lockstep sync, Asynchronous sync, Adaptive Event sync, Time Warp sync, Breathing Time Warp, Trailing State sync, Event Correlation sync and Optimistic Obsolescence-based sync.

## 7.2.1. Local Lag & Dynamic Local Lag

Local lag (LL) is used to reduce short-time inconsistencies between clients in an NVE by delaying an operation in every client a certain amount of time, called the local lag. This way, all the clients execute the same actions at the same time, despite their differences in network latency, by slowing down the responsiveness [141]. Figure 25 shows an example in a Client/Server-based solution. When the clients connect, the Server tests the latency of every client and based on the maximum value of those latencies, a fixed waiting time (local lag) is set for each client, which, when added up to the network latency of the client, will be equal to that maximum delay for all the clients. This technique can be further improved by using Dynamic local lag (DLL), in which, instead of using a fixed amount of delay, the information is delayed dynamically according to the network latency in both source and destination clients [135].

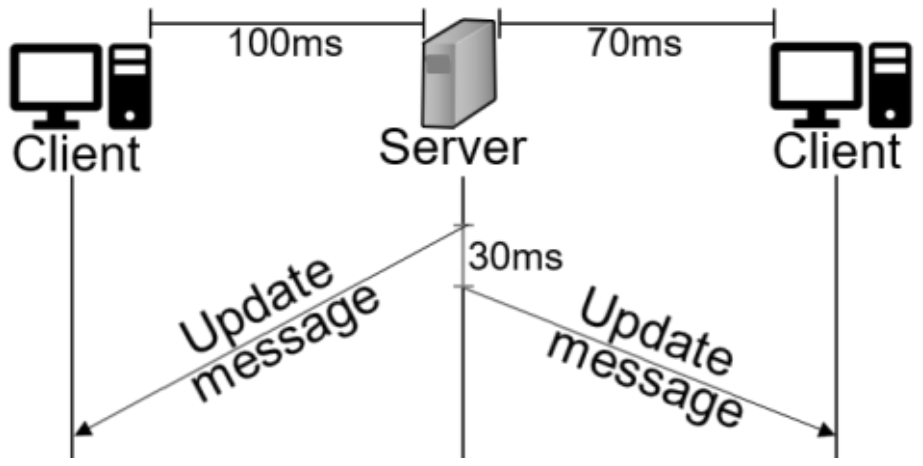

Figure 25. Local lag in a Client/Server-based solution. The update message of an event is delayed 30ms for the client with lower network delay (on the right), to match the total delay to 100ms (other client on the left).

First, the local lag is calculated at the beginning for each type of entity, according to the network latency and the responsiveness requirements of the type of entity (e.g., an entity with higher relevance in the NVE would require lower delay). Then, every time the network load changes, or the position of an entity changes, the value of the local lag is calculated and updated. Update messages are stored depending on the local lag value and sent after their waiting time. Examples of both techniques can be found in [142] (LL), and [135] (DLL).

LL and DLL have a limit, as too much lag negatively affects the responsiveness, therefore, other techniques should be used in conjunction. Also, the DLL ensures an optimum delay for every participant that also is adapted to network fluctuations, but enough throughput and processing resources are needed for the periodic calculation.

## 7.2.2. Adaptive Δ-causality

All the clients of the NVE use the same maximum end-to-end delay, as in LL, which is set dynamically according to the network latency, and determines the time limit the updates can take to reach their destination [135]. The clients will send the update messages as soon as they are generated, and, if they are received before the time limit expires, they are stored until it happens, and then, they are executed. If an update is received later, it is not executed but used for estimating values (e.g., to predict the future position of an entity). An example of NVE with Adaptive Δ-causality can be found in [135]. In Figure 26, Client 1 sends update messages to Client 2, with information of the position of its avatar. The update messages 1 and 3 are received on time, but the update message 2 (including the position $x = 5$) arrives late at Client 2. If Adaptive Δ-causality were not used, the client 2 receiving the update 200 milliseconds late would update the value of $x$ to 5, since that is the value received; or discard the message and wait for the next one. However, when using Adaptive Δ-causality, the Client 2 can use the timestamps and delays to calculate a new value instead. So, the new position is calculated with, e.g., the formula: $Ve = Vl + d \times s$, where $Ve$ is the value to be estimated, $Vl$ is the value received late, $d$ is the delay of that value and $s$ is the step (or slope) in the value between the late update and its previous one. So, in this case, it will be $x = 5 + 200ms \times (5 - 2) / 600ms = 6$.

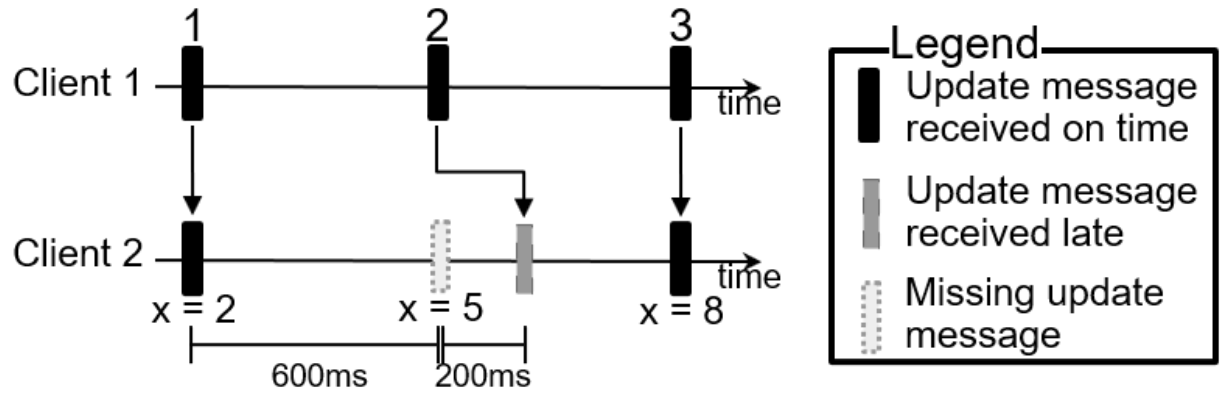

Figure 26. Adaptive Δ-causality. A late update used to estimate a position.

Adaptive Δ-causality keeps the causal relationship among messages. Thanks to this, the consistency can be maintained when the network latency is not too high. Otherwise, other techniques should be combined.

### 7.2.3. Bucket Synchronization (BS)

The main idea of BS is that messages originated by all the clients at the same time or during the same period should be processed together and at the same time in all of them. In BS, time is divided into time slots with a fixed length and a bucket is associated to each slot (called bucket period) [58]. All the update messages received by a client that were generated and transmitted by sender clients during a given period are stored by the receiver clients in the bucket corresponding to that period. At the end of each bucket interval, the receiver clients compute all the (own and received) update messages in that bucket to get their new local views of the global state of the NVE. With BS, the simulation runs in cycles of the same duration (bucket period). After each cycle, clients synchronize with the rest and then, its virtual time is increased the same amount, that is, a bucket period. However, when an update message is received after its bucket period, due to, e.g., network delay, the clients return the states to that previous bucket and repeat the execution of all its messages (rollback). When late update messages from other clients, or other rollbacks affect more than one bucket, cascade rollbacks are conducted, forcing several buckets to be reprocessed successively. An example of the use of BS can be found in MiMaze [11].

In Figure 27, the bucket ‘a’ receives the update messages 1 and 2 unordered, but all of them belonging to that same bucket, so they can be processed together without causing rollbacks nor inconsistencies. During the bucket ‘b’, update message 4 is received, but the message 3 is missing. The received message 4 is processed, therefore, causing an inconsistency. During bucket ‘c’, the late update message 3 is received, forcing a rollback, reprocessing the previous bucket ‘b’, and then, the bucket ‘c’, recovering the consistency.

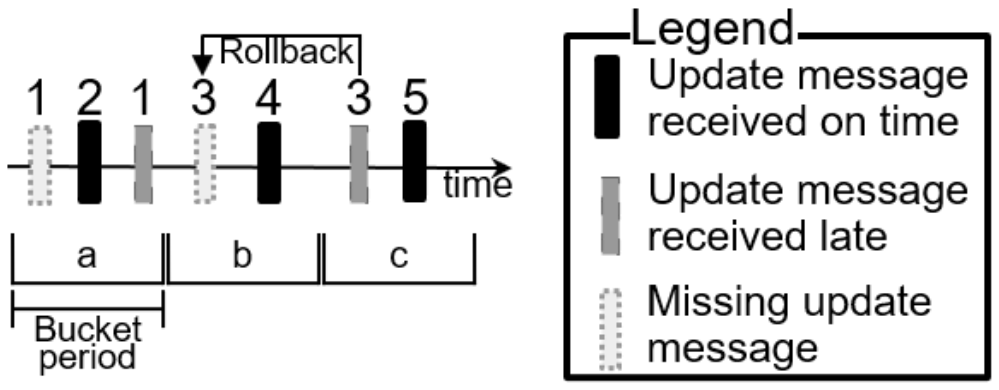

Figure 27. Example of BS with a rollback.

The main advantage of BS is its low computation overhead required, but it presents several flaws that make it impossible to be used in NVEs requiring a high level of responsiveness. These disadvantages are mainly related to the bucket period, as it should be long enough, so each client can process enough messages, but it should also be short enough to support fast and realistic interactions.

### 7.2.4. Breathing Time Buckets (BTB)

BTB is like BS, but, in this case, the length of the bucket periods is variable, and each client processes their buckets independently of the others [143]. During a bucket period, the received update messages are executed as they arrive if they do not precede another that has already been processed. Otherwise, if rollbacks are needed, the bucket duration ends prematurely, while corrections are made. With BTB, when this situation happens, control messages are interchanged between clients so that rollbacks are done locally, and no cascade rollback occurs. An example of the use of BTB can be found in [144].

In Figure 28, an example of a transmission of update messages between two clients is shown. Client 1 sends the update message 3 that arrives late (outside its corresponding bucket period) to Client 2, but it does not cause a rollback in that client because it does not break consistency (e.g., it is an independent event). Nevertheless, when Client 2 sends the update message 4 to Client 1, it arrives too late and generates a rollback on Client 1 because this update message is received after update message 5 that depends on the update message 4, and the consistency had been broken.

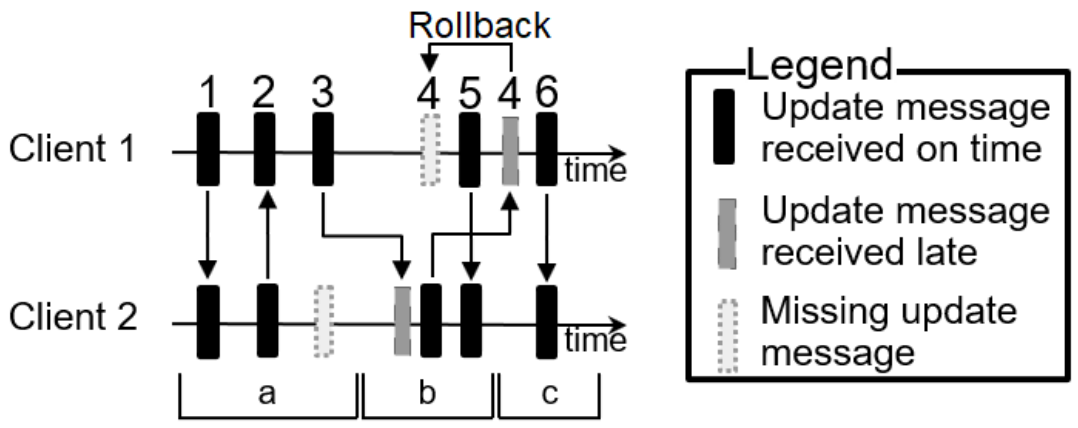

Figure 28. Example of BTB with different sized buckets.

With BTB, network usage is reduced, since most rollbacks happen locally, and responsiveness is increased. Nevertheless, the global consistency cannot be ensured, and enough messages are required in the same bucket to reduce the number of rollbacks.

## 7.2.5. Lockstep synchronization (LS)

In LS, a server manages a global time reference. For every interaction (event) in the NVE that changes any states of entities, the server stops the simulation time until all the participants update their states for those entities. Clients do not advance in time until the server notifies them to do so. Then, the simulation time is resumed. This way, a consistent NVE is achieved [26]. In Figure 29, an example of the steps followed for each interaction in a simple scenario are described. Firstly, when a user wants to interact, the associate client informs the server. Then, the server stops the global simulation time or GST (lockstep mode), notifies it to the rest of clients and sends them the event, so each client can start its computation processing. When finished, each client notifies it to the server. Only when all the clients' notifications have been received by the server, it advances the GST and moves to the next interaction or turn of the NVE. An example of the use of LS can be found in RING [26], and in [145].

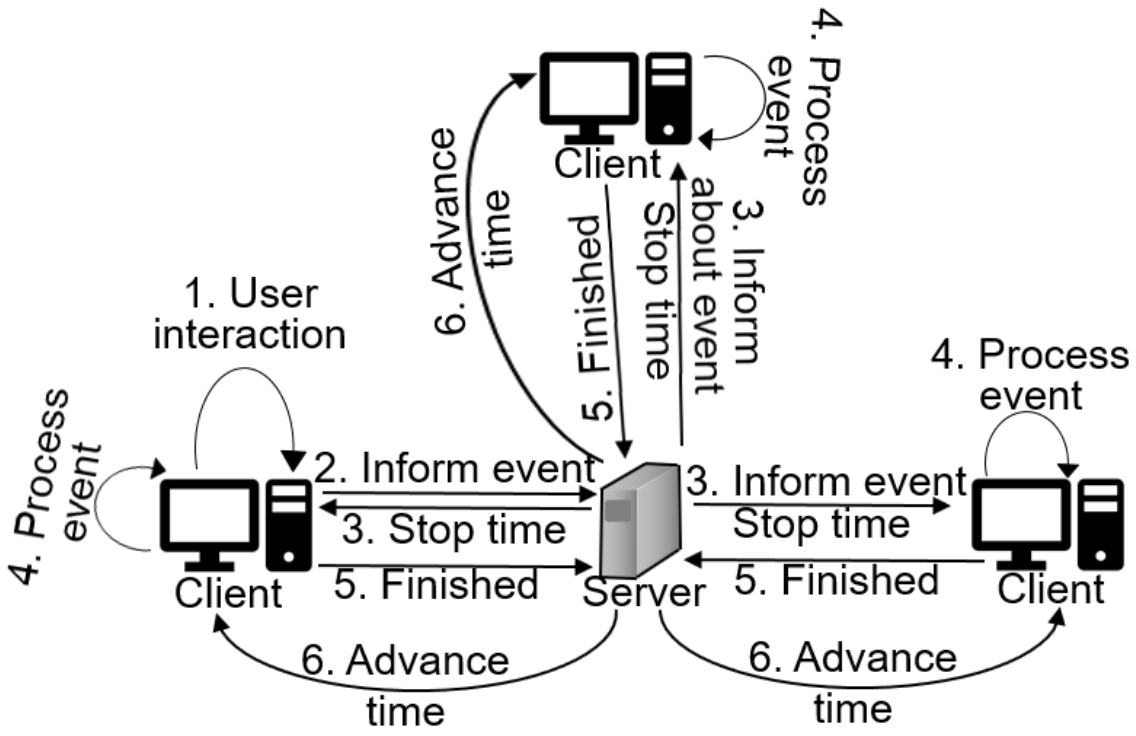

Figure 29. Steps of the Lockstep Synchronization.

With LS, the consistency is always ensured. Nevertheless, it is not recommended to be used in NVEs demanding high level of interactions and responsiveness, since clients would enter continuously in the lockstep mode, waiting for the rest of clients to notify that they finished processing the events, being very annoying for them. Furthermore, if the processing of an interaction in any client is delayed for a considerable amount of time (e.g., the end-to-end delay is high, or the update message takes too much time to be processed on one client), the server will stop the GST until that client finishes and notifies it to the server. So, the responsiveness of the NVE, and, consequently, the users' QoE, could be seriously affected.

## 7.2.6. Asynchronous Synchronization (AS)

AS is based on the previous one, but with a decentralized clock, which allows each client to advance simulation time without depending on the other clients [146]. This is achieved by only sending event update messages to the clients that are affected by those events (e.g., when a client's avatar can see another client's avatar opening a door or they are shooting each other). Each client has their own clock and perceives the same consistent world, but at different times. An example can be found in [147], where the concept of Spheres of Influence (SoI) is used. A SoI is the zone close to a clients' avatar that can be affected by it in future turns, as shown in Figure 30.

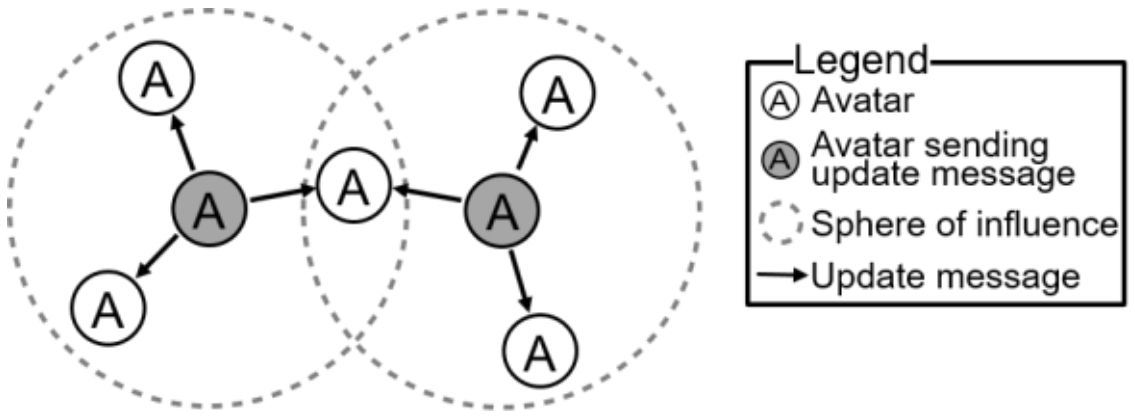

Figure 30. Example of AS. The clients inside a SoI receive the update message.

AS is resilient to game cheaters and improves the flaws of the LS, solving isolated effects of poor connection.

However, this method requires all clients to have similar network conditions, because a client that has higher latency than the rest would act as a bottleneck. Moreover, a global consistency is also hard to maintain when the number of connected clients increases.

### 7.2.7. Adaptive event synchronization (AES)

AES takes the fluctuating conditions of the network into account and combines the delay and packet loss to provide information that helps to determine the playout delay. The playout delay is a controllable delay added in each client to have the same end-to-end delay for all of them (Figure 31), attaining visualization of the events at the same time. This playout delay is calculated by selecting a main client that takes the maximum delay between the different clients, estimates the jitter based on previous known measurements, monitors the packet loss, and share all this information with the rest of clients. Then, the clients determine how much time the events will be kept in a buffer before being transmitted or executed in sync. The usage of that buffer is the main difference between AES and LL. The idea of using both the delay and loss is to solve punctual, as well as long-term inconsistencies. An example of the use of AES can be found in [148].

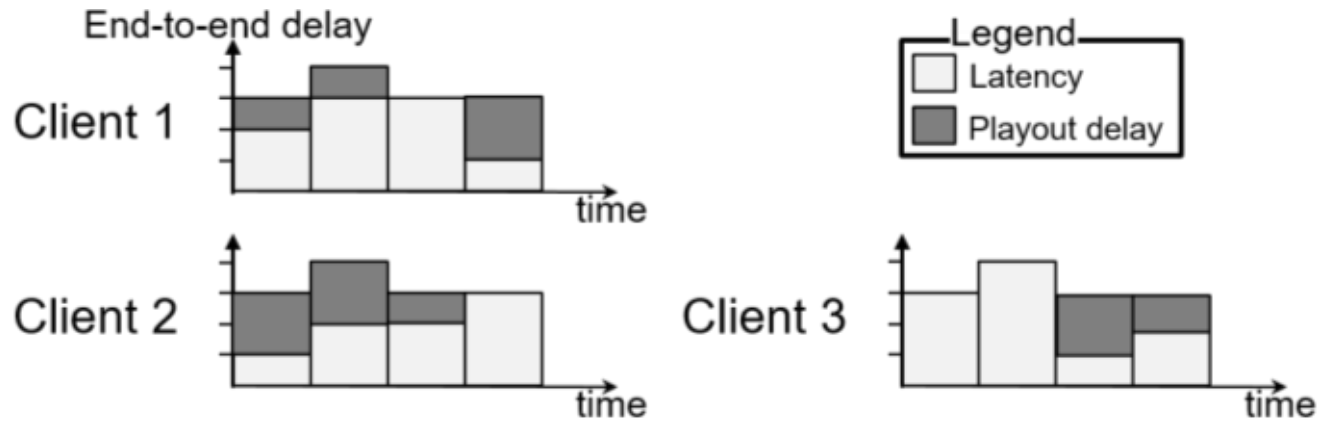

Figure 31. AES. The latency estimation from the loss and delay is used to calculate the necessary end-to-end delay to equilibrate the timings of all clients.

As network characteristics are used to establish the sync parameters, users will likely be able to interact in a consistent virtual world. Nevertheless, as one of the clients is elected to process the sync computations, it needs enough computation resources to work properly. Furthermore, AES also adds the control messages that reduce the available network throughput.

### 7.2.8. Time Warp Synchronization (TWS)

In TWS, the event-related messages interchanged in the NVE have four fields: the name of the sender, the name of the receiver, and the virtual sending and receiving timestamps [149], which are filled by the clients and used to synchronize the update messages. The sender and receiver clocks should be previously synchronized, so that the timestamps indicate the correct relationships between events. The update messages are processed by the clients as they arrive. If an update message arrives containing information of an event timestamped before the event being processed, rollbacks are made by executing the older event and then the following ones in order. As shown in Figure 32, when an event (3) is received late (a) a rollback is performed to return to the state corresponding to an execution moment before that event happened and then all the received events are processed in order (b). To avoid inconsistency problems caused by update messages that have been sent by that client before the late update message was received, that client inform the rest of the clients about the rollback and that some of its previously sent update messages could represent an incorrect state of the NVE. An example of TWS can be found in [84].

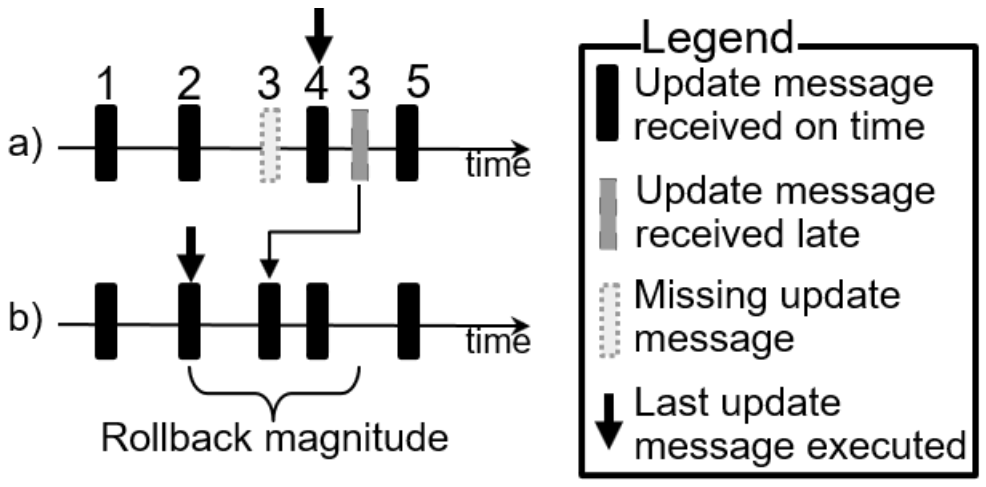

Figure 32. Rollback process in Time Warp.

TWS allows the NVE to have a high responsiveness when having enough throughput and processing resources. If these features cannot be guaranteed, TWS should be used only when the rollback processes do not occur often because they can be highly annoying for the users. The main disadvantage of TWS is the need of high memory capacity, as copies of the NVE processed messages must be stored.

### 7.2.9. Breathing Time Warp (BTW)

BTW consists in a combination of TWS and BTB, and deals with the issues they can experience [150]. First, it starts with a TWS phase, where events are treated in an optimistic way up to a chosen time delay (lookahead), performing rollbacks and informing about them if inconsistencies happen. When a specified time passes, the BTW moves into a BTB phase, until a specific number of events are processed, and rollbacks are stored to be sent later. After this, it goes back to the TWS phase and repeats the processing cycle. In Figure 33, the two delayed update messages in the TWS phase led to two rollbacks, each happening when a late update message is received. Later, in the BTB phase, the two late update messages received in bucket 'b' generate a single rollback event to the bucket 'a'. Examples of the use of BTW can be found in [150], [151].

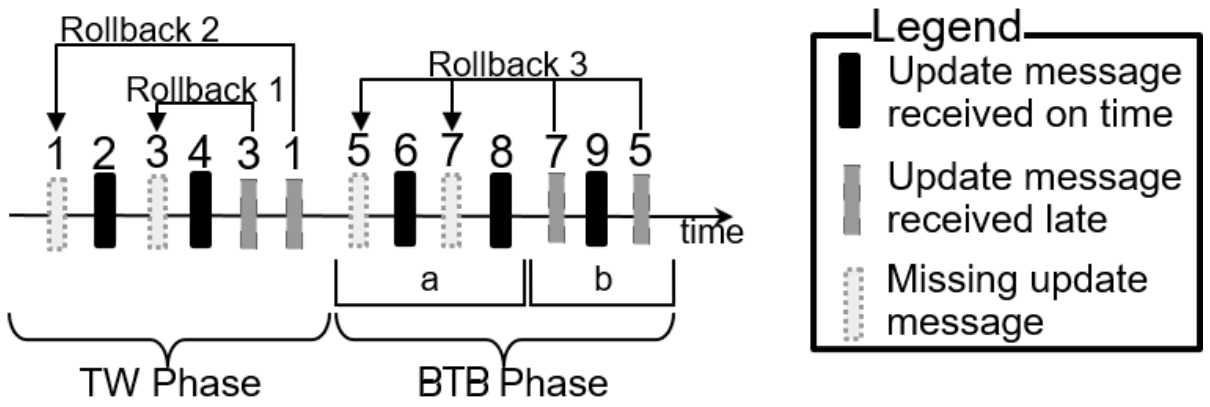

Figure 33. Example of Breathing Time Warp (TWS and BTB combination).

The problem of BTW is that the time between cycles can affect the consistency in scenarios where the number of messages varies dynamically. To solve this problem, in SafeBTW [151] the time between these two phases is changed dynamically, adapting to the network and to the number of received messages. Generally, BTW can provide a better consistency than TWS, without losing too much responsiveness like with BTB.

## 7.2.10. Trailing state synchronization (TSS)

TSS is like TWS, but in this case, a series of consistent copies of the NVE in previous execution times are stored [152]. In parallel to the main simulation of the NVE, delayed copies of itself (i.e., copies of the NVE at a different virtual time), are being executed. When a new event arrives, if it precedes the current visible state, a rollback process takes place. As there are several delayed copies of the NVE running, it is probable that one is in a state before the instant the late update message was sent, and there will not be any inconsistency in that copy of the NVE. In that case, this copy of the NVE will be turned into the main version, executing all the received updates in it in the correct order. This is represented in Figure 34, in which there is a main processing copy of the NVE and three delayed copies (*trailing state copies*) of the NVE (a). When late update message 6 arrives a kind of rollback is forced. The copy stored previously to the instant the event in the late message should happen is restored as the main processing copy and the NVEs continues reprocessing all the ordered received events since the time it was stored. Although several copies of the NVE are simultaneously running, only the main one is visible to the users. An example can be found in [152].

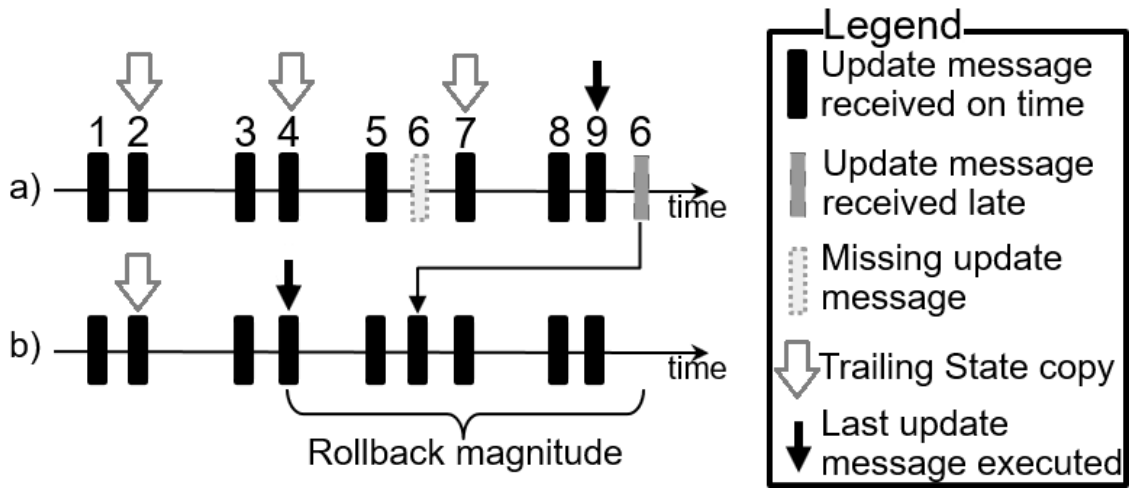

Figure 34. Example of TSS. In a), 3 TS copies are recorded holding the changes from the preceding received update messages. In b) a rollback moves the execution point of the last TS copy that did not miss any update message is.

In comparison to TWS, TSS provides better responsiveness, and the rollback process is improved, making it suitable to be used in fast-paced NVEs, such as first-person shooter (FPS) MOG. Nevertheless, it requires high memory capacity and processing resources to maintain all the needed copies of the NVE.

## 7.2.11. Event Correlation Synchronization (ECS)

ECS is like TWS and is based on event correlation algorithms. Some events are time-related between them (event-correlation) and some are not (non-event-correlation) [153]. Correlation means that one event depends on another event, so that one can happen after the other, but not vice versa. When a late update message arrives, instead of rolling back as in TWS, firstly, correlation of this event with the already processed events (stored) is checked to decide what the best choice is. If no correlation is found, the late event can be processed without rolling back and it will not lead to inconsistency. Otherwise, the rollback process will take place.

As shown in Figure 35, Client 1 sends update messages to Client 2, and the update message 2 is received after the update message 3 by the Client 2, e.g., due to fluctuations in network latency. If the events in messages 2 and 3 are correlated, so that event 2 must happen before event 3, a rollback will take place. Otherwise, update message 2 will be processed without the need of a rollback. Moreover, if the update message 3 already updated the same state that the update message 2 would (e.g., both changed an avatar's position), the late update message 2 will not be executed, as there is already a most recently updated state by the update message 3. An example of ECS can be found in [153].

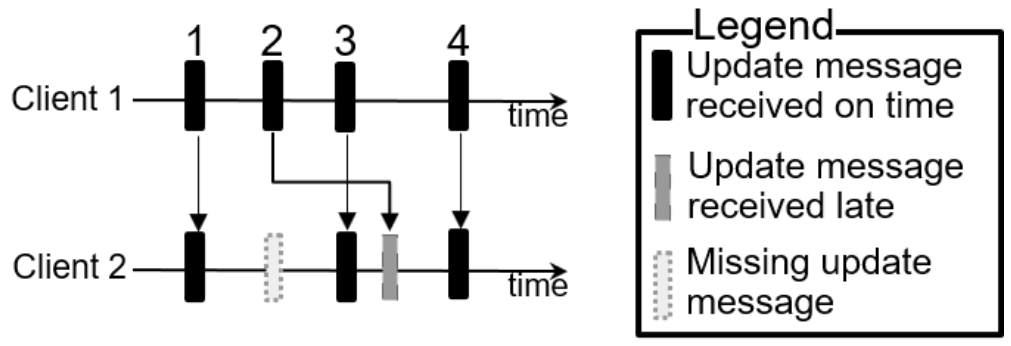

Figure 35. Example of the Event Correlation Synchronization.

ECS can reduce the number of rollbacks, improving responsiveness and interactivity and lowering the network usage and avoiding inconsistencies. Nevertheless, it has high memory capacity and processing resource requirements to do the correlation computing.

### 7.2.12. Optimistic Obsolescence-based Synchronization (OOS)

In OOS, besides applying TWS and ECS, event obsolescence is also considered. If an event arrives after it has already been overridden by a newer one, it is discarded [154], [155]. Given two update messages, *i*, and *j*, j makes *i* obsolete if processing *j* (generated after *i*) without *i* can achieve the same final state that would be reached if both events were processed in the correct order. The obsolete events are discarded, so there will not be a need to check for inconsistencies and no rollback will happen. For instance, *i* could be an update message that changed the position of an entity, while *j* could be an update message that destroyed this entity (eliminating it from the virtual world) and, therefore, the older position update becomes unnecessary. Examples of the use of OOS can be found in [156], [157].

As in TWS, in OOS a list of the processed events is also maintained, and responsiveness is improved. However, in OOS, the number of rollback actions is highly decreased, reducing the users' annoyance and improving interactivity and users' QoE. However, like in ECS, this one also requires high memory capacity and processing resources to execute the required processes.

### 7.2.13. Comparison

Table 8 presents a summary of the sync techniques presented in this section, including their advantages, disadvantages, as well as some examples of NVEs using them.

| Synchronization technique | Advantages | Disadvantages | Examples |
|---|---|---|---|
| Local lag (LL) | • Easy consistency control<br>• Reduced network usage | • Responsiveness problems<br>• Robustness problems | Khan et al. [142] |
| Dynamic local lag (DLL) | • Easy consistency control | • Control messages add to network usage | Huang et al. [135] |
| Adaptive ∆-causality | • Easy consistency and responsiveness control | • Requires low network delay | Huang et al. [135] |
| Bucket synchronization (BS) | • Low sync overhead | • Reduced Responsiveness | MiMaze [11] |
| Breathing time buckets (BTB) | • Better responsiveness<br>• Reduced network usage | • Hard to keep global consistency | Damitio et al. [144] |
| Lockstep synchronization (LS) | • Ensured consistency<br>• Easy to implement | • Not suited for real-time applications<br>• Not suitable if high responsiveness is needed | RING [26]; Chen et al. [145] |
| Asynchronous synchronization (AS) | • Resilience to cheaters<br>• Good local consistency | • Hard to keep global consistency<br>• Not suitable for fast-paced NVEs | Baughman et al. [147] |
| Adaptive event synchronization (AES) | • Good local consistency and responsiveness | • High computation resources needed<br>• Control messages increase network use | Kim et al. [148] |
| Time warp synchronization (TWS) | • Good responsiveness | • Increased network usage<br>• High memory capacity needs<br>• Rollbacks | Nguyen et al. [84] |
| Breathing time warp (BTW) | • Good consistency and responsiveness | • Complex to implement<br>• High computation resources needed | Steinman et al. [150]; SafeBTW [151] |
| Trailing state synchronization (TSS) | • Good responsiveness (better than TWS)<br>• Suited for real-time applications | • High computing resources needed<br>• High memory capacity needs<br>• Rollbacks | Cronin et al. [152] |
| Event correlation synchronization (ECS) | • Good responsiveness (better than TSS)<br>• Decrease network usage<br>• Reduced number of Rollbacks | • High computing resources needed<br>• High memory capacity needs<br>• Rollbacks | Bin Shi et al. [153] |
| Optimistic obsolescence-based synchronization | • High responsiveness and good consistency | • High computing resources needed<br>• High memory capacity needs<br>• Rollbacks | Ferretti et al. [156], [157] |

Table 8. Sync techniques in NVEs.

### 7.2.14. Future research directions

In the future, newer and more advanced solutions could apply artificial intelligence (AI) or machine learning techniques to model and predict things like the occurrence of events, the network conditions, user behavior, etc., so the synchronization can be enhanced [158]. For example, in [159], AI techniques are used to optimize the scheduling of events and performance on the synchronization of wireless sensor devices. So, it could be possible to contrive AI-based NVE solutions soon. Finally, more focus should be put on mobile devices, so that synchronization techniques perform better in them.

# 8. Computing models

Depending on the employed network architecture, in NVEs, the data, the tasks, and the computation needed by each of the interconnected nodes can be managed in different ways, optimizing the delivery of information and the performance of tasks, like, e.g., the rendering of the 3D virtual world. The rendering of the virtual world, other intensive tasks (e.g., a complex behavior of an entity), and the required storage size for the NVE can be delegated to remote nodes, which will provide the needed results and information for updating the virtual world and representing it in the client. Those helper nodes may be closer or farther to the client (e.g., in the Edge nodes or in the Cloud, respectively), in another peer or in the same house, and could be serving their features to a single user or to multiple ones (e.g., rendering frames for a group of users). These techniques also allow the NVE designers to provide a service-oriented solution, where the access to products, programs and other technologies are offered as services (e.g., a subscription to use an application temporarily) instead of the traditional on-premises approach. So, clients delegate part of (or all) their computational requirements and roles to third parties. In the Cloud Computing area, this is known as *aaS (“Something” as a Service). Examples are SaaS (Software as a Service), PaaS (Platform as a Service) and GaaS (Games as a Service). In the NVE scope, these techniques mitigate the problems that end-user lightweight clients (e.g., computers with low processing capabilities, or smartphones with lower storage size) experience. Furthermore, these techniques are still acceptable for the rest of clients if the downsides they present are not severe, allowing companies to offer this business model to all the possible clients.

In this section several techniques to manage the computing processing requirements for clients in NVEs, such as Remote Rendering, Adaptive Streaming, Foveated Imaging, Memoization, and Progressive Downloading are explained.

## 8.1. Remote Rendering

Remote Rendering is based on using other computers for rendering the contents of the NVE. The resulting audio and video streams are delivered through a network connection to the clients [79]. As users interact with the NVE, their clients send the input interactions to the renderer computer to execute the necessary processes and return the results or needed information. To allow a good level of interactivity, the latency between clients and that computer needs to be low, or the responsiveness will be impaired. This becomes a troublesome task when the quality of the images or video of the NVEs is very high, and the available throughput is very low. To tackle this problem, besides the predictive modelling and sync, Adaptive Streaming and Memoization techniques (explained later) can be used.

There are two main reasons to use this model: 1) it allows lightweight clients to still run and interact with NVEs that have high computation requirements, also saving money with the Cloud-based architecture [78]; and 2) it avoids the need to store the entire NVE program (as it runs in another computer), hence saving storage capacity on the client. There are also two additional benefits that come with Remote Rendering: 1) it is platform-independent, the NVE must only be developed for the server that is going to execute it, and the same rendered frames can be transmitted to all kinds of clients and platforms (Android, Linux, Windows, PlayStation, Xbox, etc.) without restrictions; and 2) there is only one copy of the NVE, making it easier to maintain and update the NVE.

There are many examples that employ Remote Rendering. Examples of Cloud Gaming platforms are Sora Stream [160], PlayKey [161], GeForce Now [162] and Google Stadia [163]. These platforms offer games on demand for a monthly fee, all rendered from their Cloud and brought to the clients. Parsec [164] and Steam Remote Play [165] allow the users to store the games on their own computer and stream them to other devices (e.g., a TV over LAN). MUVR [88] and CloudyGame [83] allow mobile clients (e.g., smartphones) to play games using Edge computers, which do the rendering of frames instead of a farther Cloud that would increase the delay. Finally, DROVA [166] and Vectordash [167] are distributed solutions, a.k.a. P2P Cloudless Gaming, which, instead of using a Cloud, depend on a decentralized network that balances the required rendering and computation load. This way, users can have a close available computer to manage the NVE load, instead to using a Cloud-based service, improving responsiveness.

## 8.2. Adaptive Streaming

As high-quality rendered images (frames) require a high throughput, adaptive video streaming methods can be applied to reduce the throughput usage. The quality of the rendered frames must be the optimum for the available throughput (which can change dynamically depending on the fluctuation of the network conditions) [168]. With Adaptive Streaming, when the available throughput decreases, the client receives lower quality frames instead of getting them delayed or getting disconnected from the NVE session. When the available throughput increases, higher quality frames can be rendered again and transmitted. Examples of the use of Adaptive Streaming can be found in [169]–[171].

With Adaptive Streaming, the network usage is optimized, reducing congestion and possible packet loss. Nonetheless, if the available throughput is too low, the reduced quality of the received frames can provide the users with a bad QoE. It is also important to note that these frames are generated and transmitted in real time instead of stored beforehand. Therefore, an algorithm to decide the quality of the frames to be transmitted in each moment is needed.

## 8.3. Foveated Imaging

When using adaptive streaming the users can perceive a bad QoE when the throughput is too low. To solve it, with Foveated Imaging a region of interest (RoI) is defined in the viewport of the user, so that the quality of the RoI in each transmitted frame is higher than in the rest of the frame. It allows to improve the users' QoE by reducing the quality of the parts of the rendered frame the user is not paying attention to, as shown in Figure 36.

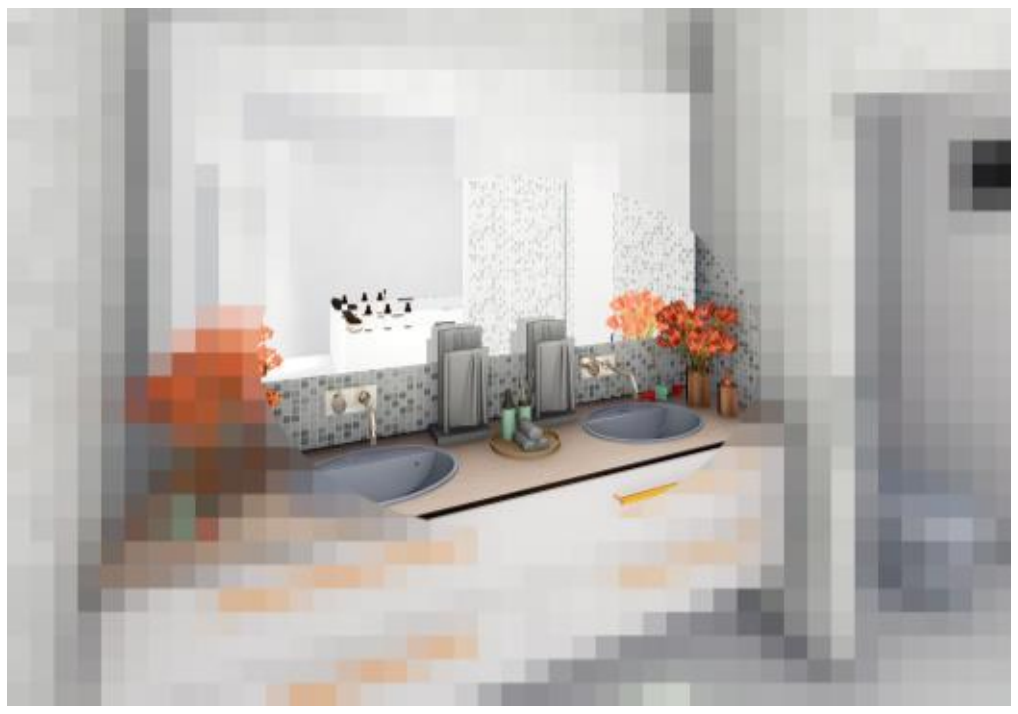

Figure 36. Foveated rendered image.

Examples of the use of Foveated Imaging can be found in [172], [173]. In the virtual reality (VR) world, this is called *foveated rendering* [172]. As users employing an HMD can move the head, and hence the viewport, freely and easier than without the HMD, the NVE should be able to send newer frames faster, updating to the new perspective and reducing the motion-to-photon latency.

The advantage of this Foveated Imaging is that the amount of transmitted information can be significantly reduced, without affecting too much to the users' QoE, improving the network usage and the experienced latency. Nonetheless, this comes with a higher computation demand.

## 8.4. Memoization

Since there can be redundancy, during the NVE session, between the content of rendered frames for different clients (e.g., users whose avatar is moving in the same environment view similar backgrounds), with the Memoization, the rendered frames are cached and reused if needed. This consists in storing rendered frames or the results of other long computations to use them again when possible [88]. This way the needed time for processing and computing resources is decreased, reducing delays, and optimizing the performance of remote rendering. An example of the use of Memoization can be found in MUVR [88], where it is used on an Edge-based architecture to reduce delays and rendering requirements between mobile clients. Every time a frame is requested from a specific position in the virtual environment (usually from the head of the user's avatar), it is checked whether there is a stored frame recorded from a similar position. If not, that frame is rendered, and then it is stored along with the position and orientation it was viewed from, for possible future uses. If cached frames exist from a similar position, the NVE combines them to generate a new frame representing what the user should view from that perspective. This is called image-based rendering (IBR) [79] and it consists in combining 2D images to simulate and render 3D points of view at different positions in the virtual world.

The NVE can also use frames that were rendered for different clients to render new ones for other clients (e.g., a cloud NVE rendering the frames from several users participating in the same session). So, a system that is rendering the NVE session for multiple clients, can take advantage of that redundancy too, reducing computing needs.

Figure 37 illustrates Memoization. In a) and b), two different points of view of the same entity are employed to store two new frames generated by 3D rendering. In c), the two frames are used to generate a new frame from a different point of view with IBR, discarding the need of 3D rendering.

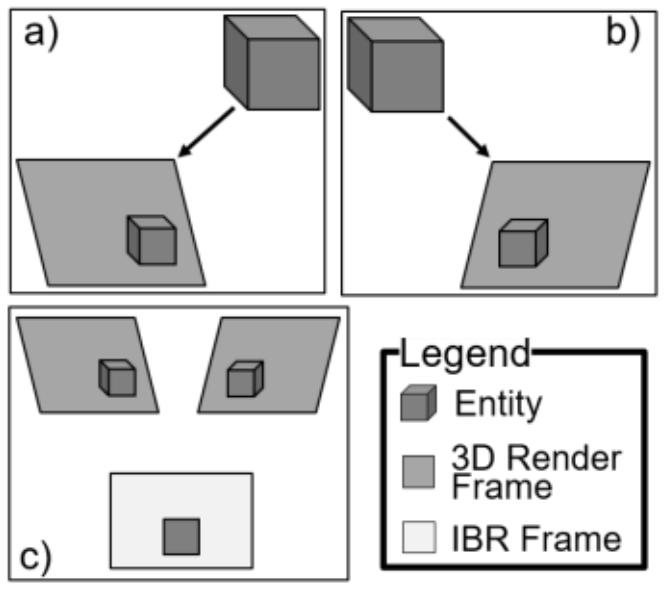

Figure 37. Two stored frames used to create a frame from a new point of view.

As IBR requires less computing power, clients with low processing capabilities can store and use cached frames to render new frames, reducing the network usage and the latency but needing more memory capacity to store them. However, the frames can become obsolete as time passes or when the virtual environment changes (e.g., entities move), forcing new frames to be rendered. Furthermore, if the number of cached frames is high, a node may choose to delete less used or older ones, to save storage size.

## 8.5. Progressive Download

The Progressive Download consists in allowing clients to execute the NVE although not all its contents are available yet [174]. This means that the NVE application can run while downloading its remaining content (Figure 38). Examples that use progressive download can be found in SuperStreamer [174] and Utomik [175].

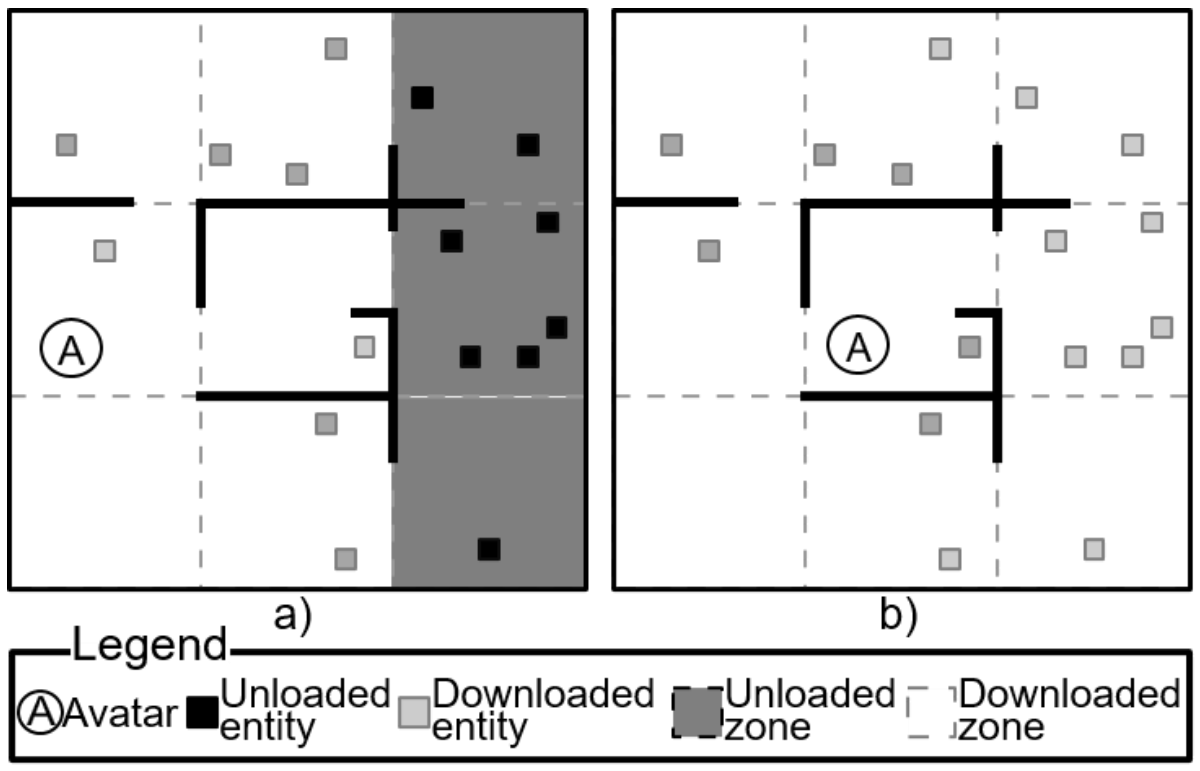

Figure 38. Progressive Download when the avatar is moving.

Thanks to Progressive Download, users start experiencing the NVE while the complete version keeps downloading. Recent videogames with high-definition graphics already surpass 100GB of storage size (e.g., Destiny 2 is 105GB [176] and Call of Duty: Modern Warfare is 175GB [177]). Furthermore, on clients with low storage capacity, Progressive Download allows to just load the information required at each moment (e.g., a virtual zone and its contents). Progressive Download is also practical for highly customizable and continuously changing NVEs, like Second Life [40], Mozilla hubs [44] and Decentraland [46], NVEs where the parts of the environment are loaded on demand, as the user requires them.

Also following Progressive Download, the Blizzard's Battle.net [178] launcher has three stages for a game download. In the first one, a predefined downloaded percentage of the game is needed to be able to run the game with reduced performance and graphics. In the second stage, after a subsequent amount has been downloaded, the game can be executed without stability issues (i.e., without the frames per second fluctuating), but with some content missing. Finally, the third stage comes when the game is fully downloaded. Moreover, the well-known videogame platforms PS4 and Xbox One have the features Play as you Download [179] and Ready to Start [180], respectively, for allowing their users to be able to play small segments of a game while the rest of it is being downloaded. Additionally, the Xbox One has another functionality called FastStart [180] that determines which resources of the game are needed before, so they can be downloaded first, allowing a user to run the game before it is fully downloaded.

## 8.6. Comparison

Table 9 summarizes the described techniques for optimizing the NVEs requirements, including their advantages and disadvantages, as well as examples of NVEs using them.

| Technique | Advantages | Disadvantages | Examples |
|---|---|---|---|
| Remote rendering | • Easy update for all clients<br>• Reduces computing and storage requirements of the end clients | • High bitrate requirements<br>• Sensible to delays | MUVR [88]; CloudyGame [83]; Sora Stream [160]; PlayKey [161]; GeForce Now [162]; Stadia [163]; Parsec [164]; Remote Play [165]; DROVA [166]; Vectordash [167] |
| Adaptive streaming | • Reduces network usage | • Can worsen the QoE | Rhee et al. [169]; Hong et al. [170]; Wang et al. [171] |
| Foveated imaging | • Reduces network usage<br>• Low impact on QoE | • Increases computation requirements | Illahi et al. [172]; Ahmadi et al. [173] |
| Memoization | • Reduced overload of the Cloud<br>• Lower requirements for clients | • Not suited for dynamic worlds<br>• Balance needed between the storage and computing | MUVR [88] |
| Progressive downloading | • Data easy to manage by the NVE owner<br>• Reduce storage needs<br>• Reduces waiting times | • Increases NVE complexity<br>• Data downloading decreases available throughput employed for interaction | SuperStreamer [174]; Utomik [175]; Second Life [40]; Mozilla hubs [44]; Decentraland [46]; Battle.net [178]; PS4 [179]; Xbox One [180] |

Table 9. Computing models in NVEs.

## 8.7. Future research directions

For the remote rendering of stereoscopic vision (e.g., VR on HMDs), there is no standardized solution for image compression yet that considers the redundancy existing between the two frames rendered for the two eyes [88]. These frames, being rendered at the same moment, are oftentimes quite similar, and the transmission of them could take advantage from a streaming technique that reduces that extra network usage for VR, this way improving the current compression solutions, like in Adaptive Streaming and Foveated Imaging. Other works that also study this redundancy between eyes and between different clients can be found in DeltaVR [181] and in Coterie [182]. Moreover, in [183], there is a study on modeling the viewpoint (or gaze) of users in VR, which could be useful to improve the performance of Foveated Imaging.

Additionally, the IBR can also be used for local rendering, to save computation resources that could be used by other techniques, and the same happens with memoization, which can be used for storing results of other computing intensive tasks, like the ones a prediction algorithm could do, and store the results to reduce the delay added by computing processes. As techniques applied specially on NVEs are yet to be found, future research in this field could focus on implementing this kind of techniques in NVEs.

Furthermore, upscaling and sharpening techniques have been used recently in videogames to improve graphic quality of the rendered images, by increasing the image resolution and detail, without much increase on the computation costs, with the most notable example being NVIDIA DLSS (Deep Learning Super Sampling) [184]. The problem of this technology is that it is applied at the rendering process, meaning that remote rendered images get upscaled before being transmitted to the client, instead of sending the downscaled version, and letting the client, or a node closer than the Cloud, to upscale the frame, improving the network usage. Overall, more techniques in these fields are expected to be developed and applied to NVEs alongside the increasing requirements of more modern NVEs.

# 9. Conclusions and Future Work

This paper is intended to serve as a starting point for future investigations in the NVE field as well as a handy tool for future NVE developments. It provides a big picture of the main technologies employed for designing NVEs. First, some NVE-related background information has been provided and very important factors to be considered in NVEs are described, such as consistency, responsiveness, concurrency, and synchronization, among some others, to guarantee an overall satisfaction of the users (i.e., a good QoE). Additionally, the problems faced when implementing NVEs have been discussed, justifying the need for techniques to solve them.

Then, an up-to-date review and compilation of the most important network architectures, computing models, data distribution models, and techniques for resource balancing, predictive modeling, and synchronization used in NVEs, has been presented. They have been revised, compared, and classified, while also mentioning the diverse fields that may boost interest to explore.

It has been shown how the different techniques manage the described important factors in an NVE, requiring a classification in different components. A novel taxonomy has been provided as an assistance tool in the study of NVE techniques, and to classify new techniques to appear in the future. This classification is based on the nature of those techniques, to make it simpler to extract the relationship between them and to choose the most appropriate ones for each NVE to be designed.

However, more research is still needed in the NVE field to improve the users' QoE, as the field is continuously growing with new and better components and techniques. In the future, along with the new technological advances, users will continue demanding even better quality and functionalities, which will pose new challenges for researchers.

As future work, authors would like to update the survey with the new NVE components or new techniques for NVEs that appear in the future. Moreover, a document with some recommendations of combinations of different techniques presented in this paper for typical examples of NVEs, based on their specific requirements, will be prepared.

# References[2]

[2]The availability of the references has been checked on 11 June 2021.

## Appendix I. Abbreviations

- 3D: Three-Dimensional.
- AES: Adaptive Event Synchronization.
- AI: Artificial Intelligence.
- AS: Asynchronous Synchronization.
- BS: Bucket Synchronization.
- BTB: Breathing Time Buckets.
- BTW: Breathing Time Warp.
- CAVE: Cave Assisted Virtual Environment.
- DB: Database.
- DCS: Distributed Control Scheme.
- DHT: Distributed Hash Tables.
- DIS: Distributed Interactive Simulation.
- DLL: Dynamic Local Lag.
- DLSS: Deep Learning Super Sampling.
- ECS: Event Correlation Synchronization.
- EWMA: Exponentially Weighted Moving Average.
- FPS: First-Person Shooter.
- GaaS: Games as a Service.
- GST: Global Simulation Time.
- HCI: Human-Computer Interaction.
- HLA: High-Level Architecture.
- HMD: Head-Mounted Display.
- HW: Hardware.
- IBR: Image-Based Rendering.
- IDES: Inter-Device Synchronization.
- IDMS: Inter-Destination Media Synchronization.
- IoT: Internet of Things.
- LAN: Local Area Network.
- LL: Local Lag.
- LS: Lockstep Synchronization.
- M/S: Master-Slave.
- MMO: Multiplayer Massively Online.
- MOG: Multiplayer Online Game.
- MU: Media Unit.

- NVE: Networked Virtual Environment.
- OOS: Optimistic Obsolescence-based Synchronization.
- OS: Operating System.
- P2P: Peer-to-Peer.
- PaaS: Platform as a Service.
- PS4: PlayStation 4.
- PVS: Potentially Visible Sets.
- QoE: Quality of Experience.
- RoI: Region of Interest.
- RTCP: RTP Control Protocol.
- RTP: Real-time Transport Protocol.
- SaaS: Software as a Service.
- SMS: Synchronization Maestro Scheme.
- SoI: Sphere of Influence.
- SP: Super Peer.
- SW: Software.
- TCP: Transmission Control Protocol.
- TSS: Trailing State Synchronization.
- TV: Television.
- TW: Time Warp.
- TWS: Time Warp Synchronization.
- UDP: User Datagram Protocol.
- UFR: Update-Free Regions.
- UTC: Coordinated Universal Time.
- VE: Virtual Environment.
- VR: Virtual Reality.